\documentclass{pasa}%

\usepackage[T1]{fontenc}
\usepackage{ae,aecompl}
\usepackage[utf8]{inputenc}

\usepackage{graphicx} 
\DeclareGraphicsExtensions{.pdf,.png,.jpg,.ps}
\graphicspath{{./plots},{./submittedPlots}}

\usepackage{xcolor}
\usepackage{amssymb}  
\usepackage{gensymb}
\usepackage[nolist]{acronym}
\usepackage{ulem}
\usepackage{bm}
\usepackage{etoolbox}
\usepackage{hyperref}

\definecolor{mustard}{rgb}{1.0, 0.86, 0.35}
\definecolor{cyan(process)}{rgb}{0.0, 0.72, 0.92}

\definecolor{ochre}{rgb}{0.8, 0.47, 0.13}

\newcommand\totalMassCatalogue{86,000,000}
\newcommand\numberOfBinaries{1,000,000}
\newcommand\numberOfDNSsWithCEEs{15,201}
\newcommand\numberOfDNSsWithCEEsWeighted{365}
\newcommand\effectivelySingle{10}

\newcommand\fractionChannelI{69\%}
\newcommand\fractionChannelII{14\%}

\newcommand\lowLimLuminosity{4.3}
\newcommand\uppLimLuminosity{5.5}
\newcommand\lowLimTemperature{3.5}
\newcommand\uppLimTemperature{4.0}

\newcommand\percentageHG{4}
\newcommand\percentageGB{7}
\newcommand\percentageCHeB{59}
\newcommand\percentageEAGB{30}

\newcommand\lowLimMass{8}
\newcommand\uppLimMass{29}

\newcommand\lowLimCoreMassFraction{0.2}
\newcommand\uppLimCoreMassFraction{0.5}

\newcommand\lowLimEbind{46.8}
\newcommand\uppLimEbind{49.5}

\newcommand\lowLimSemiMajorAxis{330}
\newcommand\uppLimSemiMajorAxis{7000000}
\newcommand\lowLimPeriastron{7}
\newcommand\uppLimPeriastron{3100}

\newcommand\lowLimTotalMass{9}
\newcommand\uppLimTotalMass{37}
\newcommand\lowLimMassRatio{0.05}
\newcommand\uppLimMassRatio{1.11}

\newcommand\howManyCircConv{82}

\newcommand\howManyEccRad{68}
\newcommand\howManyEccConv{18}
\newcommand\howManyEccBelc{40}

\newcommand\fractionGiants{0.37}
\newcommand\fractionCool{0.38}
\newcommand\fractionHot{0.25}

\newcommand\LRNevent{M101 OT2015-1}

\newcommand\AllBNSRateGpcYear{85}
\newcommand\MergingBNSRateGpcYear{60}

\usepackage{xspace}
\newcommand{\monei}{\ensuremath{m_{1}}\xspace}
\newcommand{\mtwoi}{\ensuremath{m_{2}}\xspace}
\newcommand{\ai}{\ensuremath{a}\xspace}
\newcommand{\qi}{\ensuremath{q_{\rm{ZAMS}}}\xspace}

\newcommand{\vk}{\ensuremath{v_{\rm{k}}}\xspace}
\newcommand{\thetak}{\ensuremath{{\theta}_{\rm{k}}}\xspace}

\newcommand{\ei}{\ensuremath{{e}}\xspace}

\newcommand{\kms}{\ensuremath{\,\rm{km}\,\rm{s}^{-1}}\xspace}
\newcommand{\Msun}{\ensuremath{\,\rm{M}_{\odot}}\xspace}

\newcommand{\AU}{\ensuremath{\,\mathrm{AU}}\xspace}

\acrodef{DCO}{double compact object}
\acrodefplural{DCO}[DCOs]{double compact objects}
\acrodef{GRB}{gamma--ray burst}
\acrodefplural{GRB}[GRBs]{gamma--ray bursts}
\acrodef{AIS}{adaptive importance sampling}
\acrodef{CDF}{cumulative distribution function}
\acrodefplural{CDF}[CDFs]{cumulative distribution functions}
\acrodef{HR}{Hertzsprung-Russell}

\acrodef{USSN}{ultra-stripped supernova}
\acrodefplural{USSN}[USSNe]{ultra-stripped supernovae}
\acrodefplural{CCSN}[CCSNe]{core-collapse supernovae}

\acrodef{MS}{main-sequence}
\acrodef{HG}{Hertzsprung gap}
\acrodef{HSB}{hydrogen-shell burning}
\acrodef{GB}{giant branch}
\acrodef{CHeB}{core helium burning}
\acrodef{EAGB}{early asymptotic giant branch}
\acrodef{HeMS}{helium main-sequence}
\acrodef{HeHG}{helium Hertzsprung-gap}

\acrodef{NS}{neutron star}
\acrodefplural{NS}[NSs]{neutron stars}
\acrodef{DNS}{double neutron star}
\acrodefplural{DNS}[DNSs]{double neutron stars}
\acrodef{CE}{common--envelope}
\acrodefplural{CE}[CEs]{common--envelopes }
\acrodef{CEE}{common-envelope episode}
\acrodefplural{CEE}[CEEs]{common--envelope episodes}
\acrodef{RLOF}{Roche-lobe overflow}

\acrodef{SN}{supernova}
\acrodefplural{SN}[SNe]{supernovae}
\acrodef{CCSN}{core-collapse supernova}
\acrodefplural{CCSN}[CCSNe]{core-collapse supernovae}
\acrodef{PISN}{pair-instability supernova}
\acrodefplural{PISN}[PISNe]{pair-instability supernovae}
\acrodef{O/C}{oxygen-carbon}
\acrodef{ZAMS}{zero-age main-sequence}
\acrodef{TAMS}{terminal-age main sequence}
\acrodef{HR}{Hertzsprung-Russell}

\acrodef{GSMF}{Galaxy mass function, the number density of galaxies per logarithmic mass bin}
\acrodef{MZR}{Mass-metallicity relation}
\acrodef{SFR}{star formation rate}

\acrodef{DNS}{double neutron star}
\acrodef{DCO}{double compact object}
\acrodef{NS}{neutron star}
\acrodef{BH--NS}{black hole-neutron star}
\acrodef{GRB}{gamma--ray burst}
\acrodef{RLOF}{Roche-lobe overflow}
\acrodef{CE}{common envelope}

\acrodef{GW}{gravitational wave}
\acrodef{SN}{supernova}
\acrodefplural{SN}[SNe]{supernovae}
\acrodef{ECSN}{electron-capture SN}
\acrodef{PISN}{pair-instability SN}
\acrodefplural{ECSN}[ECSNe]{electron-capture SN}
\acrodef{USSN}{ultra-stripped SN}
\acrodefplural{USSN}[USSNe]{ultra-stripped SN}
\acrodef{CCSN}{core-collapse SN}
\acrodefplural{CCSN}[CCSNe]{core-collapse SN}
\acrodef{COMPAS}{
Compact Object Mergers: Population Astrophysics and Statistics}
\acrodef{MSSFR}{metallicity-specific star formation rate}

\graphicspath{ {plots/}, {updatedPlots/}}

\title[CEEs that lead to DNS formation]{Common--Envelope Episodes that lead to Double Neutron Star formation}

\author[Vigna-G\'omez et al.]{Alejandro Vigna-G\'omez$^{1,2,3,4,\star}$, Morgan MacLeod$^5$, Coenraad J. Neijssel$^{2,3,4}$, Floor S. Broekgaarden$^{5,3,4}$, Stephen Justham$^{6,7,8}$, George Howitt$^{9,4}$, Selma E. de Mink$^{5,8}$, Serena Vinciguerra$^{10,3,4}$, Ilya Mandel$^{3,4,2}$
\affil{$^1$ DARK, Niels Bohr Institute, University of Copenhagen, Blegdamsvej 17, 2100, Copenhagen, Denmark}
\affil{$^2$ Birmingham Institute for Gravitational Wave Astronomy and School of Physics and Astronomy, University of Birmingham, Birmingham, B15 2TT, United Kingdom}%
\affil{$^3$ Monash Centre for Astrophysics, School of Physics and Astronomy, Monash University, Clayton, Victoria 3800, Australia}
\affil{$^4$ The ARC Center of Excellence for Gravitational Wave Discovery -- OzGrav}
\affil{$^5$ Harvard-Smithsonian Center for Astrophysics, 60 Garden Street, Cambridge, MA, 02138, USA}
\affil{$^6$ School of Astronomy \& Space Science, University of the Chinese Academy of Sciences, Beijing 100012, China}
\affil{$^7$ National Astronomical Observatories, Chinese Academy of Sciences, Beijing 100012, China}
\affil{$^8$ Anton Pannekoek Institute for Astronomy, University of Amsterdam, Postbus 94249, 1090 GE Amsterdam, The Netherlands}
\affil{$^9$ School of Physics, University of Melbourne, Parkville, Victoria, 3010, Australia}
\affil{$^{10}$ Max Planck Institute for Gravitational Physics (Albert Einstein Institute), D-30167 Hannover, Germany}
\affil{$^\star$ Email: avignagomez@nbi.ku.dk}
}%

\jid{PASA}
\doi{10.1017/pas.\the\year.xxx}
\jyear{\the\year}

\usepackage{aas_macros}
\usepackage{hyperref} 
\hypersetup{colorlinks,citecolor=blue,linkcolor=blue,urlcolor=blue}

\hypersetup{draft}

\begin{document}

\begin{frontmatter}
\maketitle

\begin{abstract}
Close double neutron stars have been observed as Galactic radio pulsars, while their mergers have been detected as gamma--ray bursts and gravitational--wave sources. They are believed to have experienced at least one common--envelope episode during their evolution prior to double neutron star formation. In the last decades there have been numerous efforts to understand the details of the common-envelope phase, but its computational modelling remains challenging. We present and discuss the properties of the donor and the binary at the onset of the Roche-lobe overflow leading to these common--envelope episodes as predicted by rapid binary population synthesis models. These properties can be used as initial conditions for detailed simulations of the common--envelope phase. There are three distinctive populations, classified by the evolutionary stage of the donor at the moment of the onset of the Roche-lobe overflow: giant donors with fully--convective envelopes, cool donors with partially--convective envelopes, and hot donors with radiative envelopes. We also estimate that, for standard assumptions, tides would not circularise a large fraction of these systems by the onset of Roche-lobe overflow. This makes the study and understanding of eccentric mass-transferring systems relevant for double neutron star populations.
\end{abstract}

\end{frontmatter}

\section{INTRODUCTION }
\label{sec:intro}

A dynamically-unstable mass transfer episode initiated by a post-main-sequence donor is likely to lead to a \ac{CEE}, in which one star engulfs its companion and the binary spiral closer under the influence of drag forces \citep{paczynski1976common}. \acp{CEE} are proposed as a solution to the problem of how initially wide binaries, whose component stars may expand by tens to thousands of solar radii during their lifetime, become close binaries at later stages of evolution \citep{Heuvel1976closeBinaries}. Most evolutionary pathways leading to close compact binaries are expected to involve at least one \ac{CEE} \citep{ivanova2013common}.

While \acp{CEE} are frequently invoked as a fundamental part of binary evolution, the detailed physics remain poorly understood \citep{paczynski1976common,Iben1993CEE,Ivanova2013}. There have been efforts in modelling and understanding the phase through hydrodynamic simulations, using Eulerian adaptive mesh refinement \citep{Sandquist1998CEE,Ricker2008CEEs,Passy2012CEE,Ricker2012CEEs,MacLeod2015CEE,Staff2016,MacLeod2017CEE,Iaconi2017I,Iaconi2018MNRASI,Chamandy2018CEEs,De:2019,Li2019arXiv191206864L,LopezCamara2019,Shiber2019}, moving meshes \citep{Ohlmann2016CEEs,Ohlmann2017CEEs,Prust2019}, smoothed particle \citep{Rasio1996CEE,Lombardi2006,NandezIvanova2015,NandezIvanova2016,Passy2012CEE,Ivanova2016MNRAS,Reichardt2019CEE}, particle-in-cell \citep{LivioSoker1988} and general relativistic \citep{Cruz-OsorioRezzolla2020} methods. Other approaches pursue detailed stellar modeling \citep{Dewi2000CEE,Kruckow:2016tti,Clayton2017CEE,Fragos2019CEEs,Klencki2020bindingEnergy} or binary population synthesis \citep[e.g.][]{Tauris1996popSynth,nelemans2000reconstructing,dewi2006double,andrews2015evolutionary,Kruckow2018ComBinE,VignaGomez2018DNSs}. There is currently no consensus on a thorough understanding of \acp{CEE} on all the relevant spatial and time scales.

Recent rapid population synthesis studies of \ac{DNS} populations have been partially  motivated by the development of gravitational-wave astronomy.  Software tools such as \texttt{StarTrack} \citep{dominik2012double,belczynski2018GW170817,Chruslinska2017,chruslinska2018double}, MOBSE \citep{GiacobboMapelli2018,GiacobboMapelli2019ECSN,GiacobboMapelli2019,GiacobboMapelli2020}, \textsc{ComBinE} \citep{Kruckow2018ComBinE} and COMPAS \citep{VignaGomez2018DNSs,Chattopadhyay2019DNSs} have been used to explore synthetic \ac{DNS} populations in detail. Most of those studies focus on predicting or matching the observed \ac{DNS} merger rate, either by investigating different parameterisations of the physics or varying the parameters within the models. In particular, all of the aforementioned population synthesis codes follow a similar simplified treatment of the \ac{CE} phase.

In this paper, we present a detailed analysis of the \acp{CEE} that most merging \acp{DNS} are believed to experience at some point during their formation \citep{bhattacharya1991formation,belczynski2002comprehensive,Ivanova2003DNSs,dewi2003late, 2005MNRAS.363L..71D,tauris2006formation,andrews2015evolutionary, tauris2017formation,belczynski2018GW170817,Kruckow2018ComBinE,VignaGomez2018DNSs}. \cite{dominik2012double} previously used rapid population synthesis to study the relationship between  \acp{CEE} and \ac{DNS} merger rates. In this study, we focus our attention on the properties at the onset of the \ac{RLOF} episode leading to the \ac{CEE}.  We consider both long-period non-merging \acp{DNS} as well as short-period merging \acp{DNS}. We propose these distributions of binary properties as initial conditions for detailed studies of \acp{CEE}.  We provide the results of this study in the form of a publicly available catalogue\footnote{The database and resources can be found on \href{https://zenodo.org/record/3593843}{https://zenodo.org/record/3593843} \citep{VignaGomez2019CEEsDataset}}.

We examine the properties of binaries unaffected by external dynamical interactions that experience \acp{CEE} on their way to forming \ac{DNS} systems.   We briefly discuss \acp{CEE} leading to \ac{DNS} formation in the context of generating some of the brightest luminous red novae, which may be signatures of common-envelope ejections \citep{Ivanova2013,MacLeod2017LRN,Blagorodnova2017LRN,Pastorello2019LRNe,Howitt2019LRNe}, and Be X-ray binaries \citep{VinciguerraNeijssel2020}.

This paper is structured in the following way. Section \ref{sec:methods} describes the initial distributions and relevant physical parameterisations used in rapid population synthesis. Section \ref{sec:results} presents the results of our study, particularly \ac{HR} diagrams displaying different properties of the systems, as well as their distributions. Section \ref{sec:discussion} discusses the results and some of the caveats. Finally, section \ref{sec:summaryAndConclusions} summarises and presents the conclusions of this work.

\section{Population Synthesis Model}
\label{sec:methods}
We characterise \acp{CEE} with the rapid population synthesis element of the COMPAS suite\footnote{\href{https://compas.science/}{https://compas.science/}} \citep{stevenson2017formation,barrett2017accuracy,VignaGomez2018DNSs,Neijssel2019}.
Rapid population synthesis relies on simplified methods and parameterisations in order to simulate a single binary from the \ac{ZAMS} until stellar merger, binary disruption or \ac{DCO} formation. This approach relies on sub-second evolution of a single binary in order to generate a large population within hours using a single processor.

In COMPAS, an initial binary is defined as a gravitationally-bound system completely specified by its metallicity, component masses, separation and eccentricity at the \ac{ZAMS}. We assume that our binaries have solar metallicity $Z=Z_{\rm{\odot}}=0.0142$ \citep{asplund2009chemical}. The mass of the primary ($m_{\rm{1}}$), i.e. the more massive star in the binary at birth, is drawn from the initial mass function $dN/dm_1 \propto m_{1}^{-2.3}$ \citep{kroupa2001variation} sampled between $5 \leq m_1/\rm{M_{\odot}} \leq 100$. The mass of the secondary ($m_2$) is obtained by drawning from a flat distribution in mass ratio ($q_{\rm{ZAMS}}=m_2/m_1$) in the form $dN/dq \propto 1$ with $0.1 < q_{\rm{ZAMS}} \leq 1$ \citep{sana2012binary}. The initial separation is drawn from a flat-in-the-log distribution in the form $dN/da \propto a^{-1}$ with $0.01 < a_{\rm{ZAMS}}/\rm{AU} < 1000$ \citep{opik1924statistical}. We assume that all our binaries have zero eccentricity at formation (the validity of this assumption is discussed in Section \ref{subsubsection:eccDist}).

\subsection{Adaptive Importance Sampling}
\label{sec:AIS}
COMPAS originally relied on Monte Carlo sampling from the birth distributions described above. However, this becomes computationally expensive when studying rare events.

In order to efficiently sample the parameter space leading to \ac{DNS} formation, we adopt \texttt{STROOPWAFEL} as implemented in COMPAS \citep{broekgaarden2019stroopwafel}. 
\texttt{STROOPWAFEL} is an \ac{AIS} algorithm designed to improve the efficiency of sampling of unusual astrophysical events. 
The use of \ac{AIS} increases the fraction of \acp{DNS} per number of binaries simulated by $\sim 2$ orders of magnitude with respect to regular Monte Carlo sampling. After sampling from a distribution designed to increase \ac{DNS} yield, the binaries are re-weighted by the ratio of the desired probability distribution of initial conditions to the actual sampling probability distribution.

We use bootstrapping to estimate the sampling uncertainty. We randomly re-sample each model population with replacement in order to generate a bootstrapped distribution. We perform this process $N=100$ times to get a 10\% accuracy of the bootstrapped standard deviation. We calculate and report the standard deviation of the bootstrapped distributions as $1\sigma$ error bars.

\subsection{Underlying Physics}\label{subsec:underlyingPhysics}
We mostly follow the physical model as presented in \cite{VignaGomez2018DNSs}.
However we highlight key aspects of the model that are particularly relevant for this work, along with a few non-trivial changes to the code. See also Appendix \ref{app:popSynth} for a summary and additional details on the setup.
\begin{enumerate}
    \item We approximate the Roche-lobe radius following the fitting formula provided by \cite{Eggleton1983rocheLobe} in the form:
    \begin{equation}\label{eq:RocheLobeRadius}
	    \dfrac{R_{\rm{RL}}}{a_{\rm{p}}}= \dfrac{0.49q_{\rm{RL}}^{2/3}}{0.6q_{\rm{RL}}^{2/3}+ \rm{ln}(1+q_{\rm{RL}}^{1/3})}, 0<q_{\rm{RL}}<\infty,
    \end{equation}
     where $R_{\rm{RL}}$ is the effective Roche-lobe radius of the donor, $a_{\rm{p}}=a(1-e)$ is the periastron, $a$ and $e$ are the semi-major axis and eccentricity respectively, $q_{\rm{RL}}$ is the mass ratio; $q_{\rm{RL}}=m_{\rm{donor}}/m_{\rm{comp}}$, with $m_{\rm{donor}}$ and $m_{\rm{comp}}$ being the mass of the donor and companion star, respectively. \ac{RLOF} will occur once $R_{\rm{donor}} \geq R_{\rm{RL}}$, where $R_{\rm{donor}}$ is the radius of the donor.
    \item We use the properties of the system at the onset of \ac{RLOF} in order to determine whether the mass transfer episode leads to a \ac{CEE}. Dynamical stability is determined by comparing the response of the radius of the donor to (adiabatic) mass loss to the response of the Roche-lobe radius to mass transfer. This is done using the mass-radius exponent
    \begin{equation}\label{eq:zeta}
    \zeta_{\rm{i}}=\dfrac{d\log R_{\rm{i}}}{d\log m_{\rm{donor}}},
    \end{equation}
    where the subscript ``i" represents either the mass-radius exponent for the donor ($\zeta_{\rm{donor}}$) or for the Roche lobe ($\zeta_{\rm{RL}}$). We assume that
    \begin{equation}\label{eq:stability}
    \zeta_{\rm{donor}} < \zeta_{\rm{RL}}
    \end{equation}
    leads to a \ac{CEE}. Inspired by \citet{Ge:2015}, for \ac{MS} donors, we assume $\zeta_{\rm{donor}}=2.0$; for \ac{HG} donors, we assume $\zeta_{\rm{donor}}=6.5$. For post-helium-ignition phases in which the donor still has a hydrogen envelope, we follow \cite{soberman1997stability}. All mass transfer episodes from stripped post-helium-ignition stars, i.e. case BB mass transfer \citep{Delgado1981caseBB,Dewi2002,dewi2003late} onto a \ac{NS} are assumed to be dynamically stable. For more details and discussion, see \cite{tauris2015ultra} and \cite{VignaGomez2018DNSs}.
    \item We deviate from \citet{stevenson2017formation} and \citet{VignaGomez2018DNSs} by allowing \ac{MS} accretors to survive a \ac{CEE}. Previously, any \ac{MS} accretor was mistakenly assumed to imminently lead to a stellar merger. We now treat \ac{MS} accretors just like any other stellar type. This does not have any effect on the COMPAS \ac{DNS} population, as there are no dynamically unstable mass transfer phases with \ac{MS} accretors leading to \ac{DNS} formation (see discussion on formation history in Section \ref{subsec:formation}).
    \item We follow \citet{de1990common} in the parameterization of the binding energy ($E_{\rm{bind}}$) of the donor star's envelope ($m_{\rm{donor,env}}$) given as:
    \begin{equation}\label{eq:eBind}
        E_{\rm{bind}}=\dfrac{-Gm_{\rm{donor}}m_{\rm{donor,env}}}{\lambda R_{\rm{donor}}},
    \end{equation}
    where $G$ is the gravitational constant and $\lambda$ is a numerical factor that parameterises the binding energy.
    \item For the value of the $\lambda$ parameter, we follow the fitting formulae from detailed stellar models as calculated by \citet{xu2010binding,XuLi2010Erratum}. This $\lambda$, originally referred to as $\lambda_{\rm{b}}$, includes internal energy and is implemented in the same way as  $\lambda_{\rm{Nanjing}}$ in \texttt{StarTrack} \citep{dominik2012double}. 
    Additionally, we fixed a bug which underestimated the binding energy of the envelope. We discuss the effect this has on the \ac{DNS} population in Section \ref{subsec:comparison}.
    \item We use the $\alpha\lambda$-formalism \citep{webbink1984double,de1990common} to determine the post-\ac{CEE} orbit, with $\alpha=1$ in all of our \acp{CEE}.
    \item We use the \citet{fryer2012compact} \textit{delayed} supernova remnant mass prescription,  which was the preferred model from \cite{VignaGomez2018DNSs}. This prescription allows for a continuous (gravitational) remnant mass distribution between \acp{NS} and black-holes, with a transition point at $2.5\ \rm{M_{\odot}}$.
    \item The remnant mass of \acp{NS} with large baryonic mass previously only accounted for neutrino mass loss instead of an actual equation of state. This has now been corrected. This only affects \acp{NS} with remnant masses larger than $\approx 2.15\ M_{\rm{\odot}}$.
\end{enumerate}

\subsection{Tidal Timescales}\label{subsec:tides}
Mass transfer episodes occur in close binaries that experienced tidal interactions. 
The details of these tidal interactions are sensitive to the properties of the binary and the structure of the envelope of the tidally distorted stars, either radiative or convective.

The equilibrium tide refers to viscous dissipation in a star that is only weakly perturbed away from the shape that it would have in equilibrium \citep{Zahn1977Tides}.  Meanwhile, the dynamical tide \citep{Zahn1975Tides} refers to the excitation of multiple internal modes of a star in a time-varying gravitational potential; when these oscillatory modes are damped, orbital energy is lost to thermal energy \citep{Eggleton:1998,MoeKratter2018}. Tidal evolution tends to align and synchronise the component spins with the direction of the orbital angular momentum vector and circularise the binary \citep{Counselman1973Tides,Zahn2008Tides}.

There are numerous uncertainties in tidal evolution. For example, the role of eccentricity is an active field of research. Heartbeat stars are eccentric binaries with close periastron passage which experience tidal excitation of different oscillatory modes (see \citealt{Shporer_2016} and references therein). Eccentric systems may also experience resonance locking, which occurs when a particular tidal harmonic resonates with a stellar oscillation mode; this enhances the efficiency of tidal dissipation \citep{WitteSavonije1999b,WitteSavonije1999}. The high-eccentricity regime, previously studied in the parabolic (and chaotic) limit \citep{Mardling1995a,Mardling1995b}, has recently being revisited in the context of both dynamical \citep{VickLai2018dynamicalTides} and equilibrium tides \citep{VickLai2019convectiveTides}. There are uncertainties in the low-eccentricity regime: for example, there is a range of parameterisations for the equilibrium tide, such as the weak friction approximation, turbulent viscosity and fast tides \citep{Zahn2008Tides}.

Here, we make several simplifying approximations for the synchronisation and circularisation timescales, $\tau_{\rm{sync}}$ and $\tau_{\rm{circ}}$ respectively, in order to parameterise the tidal evolution of the system.

We assume that the equilibrium tide operates on all stars with a convective envelope, regardless of the binary eccentricity.  We use the equilibrium tide description in the weak friction model as described by \cite{Hut1981Tides} and implemented by \citet{hurley2002evolution}, although this may not be accurate for high-eccentricity systems (but see \citealt{VickLai2019convectiveTides}). Since the equilibrium tide is generally a more efficient energy transport/dissipation mechanism than the dynamical tide for stars with convective envelopes, we ignore the contribution of the latter.  Our equilibrium tide model is summarised in Section \ref{subsubssec:Hut}.

We apply the dynamical tide only to stars with a radiative envelope.   In Section \ref{subsubsec:Zahn} we present our implementation of the dynamical tide following \cite{Zahn1977Tides}, as used in \citet{hurley2002evolution}.

\subsubsection{The equilibrium tide for stars with convective envelopes}
\label{subsubssec:Hut}
Under the equilibrium tide, the synchronisation and circularisation evolution equations for tides acting on a star of mass $m_{\rm{tide}}$ from a companion star with mass $m_{\rm{comp}}$  are
\begin{equation}\label{eq:syncHut}
    \begin{split}
    \dfrac{d\Omega{\rm{spin}}}{dt}= &\ 3\bigg( \dfrac{k}{\tau_{\rm{tide}}} \bigg) \dfrac{q^2}{r_{\rm{g}}^2} \bigg( \dfrac{R_{\rm{tide}}}{a} \bigg)^6 \dfrac{\Omega_{\rm{orb}}}{(1-e^2)^6} \\
    &\times \bigg[ f_2(e^2)-(1-e^2)^{3/2}f_5(e^2)\dfrac{\Omega_{\rm{spin}}}{\Omega_{\rm{orb}}} \bigg]
    \end{split}
\end{equation}
and
\begin{equation}\label{eq:circHut}
    \begin{split}
    \dfrac{de}{dt}= &-27\bigg( \dfrac{k}{\tau_{\rm{tide}}} \bigg) q(1+q) \bigg( \dfrac{R_{\rm{tide}}}{a} \bigg)^8 \dfrac{e}{(1-e^2)^{13/2}} \\
    &\times \bigg[ f_3(e^2)-\dfrac{11}{18}(1-e^2)^{3/2}f_4(e^2)\dfrac{\Omega_{\rm{spin}}}{\Omega_{\rm{orb}}} \bigg],
    \end{split}
\end{equation}
where $f_{n}(e^2)$ are polynomial expressions given by \citet{Hut1981Tides}. The structure of the tidally deformed star is parameterised by $k$, which is the apsidal motion constant \citep{Lecar1976Tides} and the intrinsic tidal timescale ($\tau_{\rm{tide}}$), usually associated with viscous dissipation \citep{Zahn1977Tides}. We follow \citet{hurley2002evolution} in the calculation of the $(k/\tau_{\rm{tide}})$ factor, which depends on the evolutionary stage and structure of the star. The mass ratio is defined as $q=m_{\rm{comp}}/m_{\rm{tide}}=1/q_{\rm{RL}}$ and the gyration radius as $r_{\rm{g}}=\sqrt{I_{\rm{tide}}/(m_{\rm{tide}}R_{\rm{tide}}^2)}$, where $I_{\rm{tide}}$ and $R_{\rm{tide}}$ are the moment of inertia and the radius of the tidally deformed star, respectively. The mean orbital velocity and the donor spin angular velocity are denoted by $\Omega_{\rm{orb}}$ and $\Omega_{\rm{spin}}$, respectively.

Given that $a>R_{\rm{tide}}$, for a non-synchronous eccentric binary we expect synchronisation to be faster than circularisation. If we assume that the system is synchronous ($\Omega_{\rm{orb}}=\Omega_{\rm{spin}}$), we simplify Equation \eqref{eq:circHut} and estimate the circularisation timescale as  
\begin{equation}\label{eq:circConv}
    \begin{split}
    \tau_{\rm{circ}} = &\ -\dfrac{e}{de/dt} \\
    = &\ \bigg\{ 27\bigg( \dfrac{k}{\tau_{\rm{tide}}} \bigg) q(1+q) \bigg( \dfrac{R_{\rm{tide}}}{a} \bigg)^8 \dfrac{1}{(1-e^2)^{13/2}} \\
    &\ \times \bigg[ f_3(e^2)-\dfrac{11}{18}(1-e^2)^{3/2}f_4(e^2) \bigg] \bigg\}^{-1}.
    \end{split}
\end{equation}

 \subsubsection{The dynamical tide for stars with radiative envelopes}
 \label{subsubsec:Zahn}
Following the derivation by \citet{Zahn1977Tides} we can write the synchronisation and circularisation timescales for the dynamical tide as
\begin{equation}\label{eq:syncRad}
    \begin{split}
    \tau_{\rm{sync}} = &\ 52^{-5/3} \bigg( \dfrac{R_{\rm{tide}}^3}{Gm_{\rm{tide}}} \bigg)^{1/2} \dfrac{r_{\rm{g}}^2}{q^2} (1+q)^{-5/6} \\ &\ \times E_{2}^{-1} \bigg( \dfrac{D}{R_{\rm{tide}}} \bigg)^{17/2}
    \end{split}
\end{equation}
and
\begin{equation}\label{eq:circRad}
    \tau_{\rm{circ}} = \dfrac{2}{21} \bigg( \dfrac{R_{\rm{tide}}^3}{Gm_{\rm{tide}}} \bigg)^{1/2} \dfrac{(1+q)^{-11/6}}{q} E_{2}^{-1} \bigg( \dfrac{D}{R_{\rm{tide}}} \bigg)^{21/2},
\end{equation}
where $E_2=1.592\times 10^{-9} (M/\rm{M_{\odot}})^{2.84}$ is a second-order tidal coefficient as fitted by \citet{hurley2002evolution} from the values given by \citet{Zahn1975Tides}, under the assumption (violated for some of the systems we consider) that close binaries are nearly circular. 
For the dynamical tide, we set the tidal separation (D) to be the semilatus rectum $D=a(1-e^2)$. This corresponds to the conservation of orbital angular momentum $J_{\rm{orb}} \propto \sqrt{a(1-e^2)}$.  This assumption may lead us to underestimate the circularisation timescale for stars with radiative envelopes in highly eccentric orbits.

The dynamical tide is much less efficient than the equilibrium tide for virtually all binaries; therefore, we ignore the contribution of dynamical tides for convective-envelope stars, even though they are active along with equilibrium tides.

Given the uncertainties in tidal circularisation efficiency, we do not include tides in dynamical binary evolution.  Instead, we evolve binaries without the impact of tides, then estimate whether tides would have been able to circularise the binary prior to the onset of \ac{RLOF} leading to a \ac{CEE} as described below.

\subsubsection{Radial expansion timescale}\label{subsec:radialExpansion}

The strong dependence of the tidal timescales on $R_{\rm{tide}}/a$ means that tides only become efficient when the star expands to within a factor of a few of the binary separation. Therefore, the rate of expansion of the star, which depends on the stage of stellar evolution, plays a key role in determining the efficiency of circularisation: the binary can circularise only if the circularisation timescale of an eccentric binary is shorter than the star's radial expansion timescale.  We define this radial expansion timescale as the radial e-folding time $\tau_{\rm{radial}} \equiv dt/d\log R$.
This is computed by evaluating the local derivatives within the fitting formulae of \citet{hurley2000comprehensive} to the detailed stellar models from \citet{pols1998stellar}. 

\subsubsection{Uncertainties in timescales}
The timescales defined here, rather than fully accurate descriptions of tidal evolution, are used as order of magnitude estimates to analyse the overall properties of the population. 
Tidal timescales have significant uncertainties, including in the treatment of the dominant dissipation mechanism (e.g. weak friction approximation, turbulent convection, fast tides) and their  parameterisation \citep{Zahn2008Tides} and implementation \citep{hurley2002evolution}.
\cite{Siess2013} noted the problem with the $E_2$ fit being commonly misused, both via interpolation and extrapolation of stars above $20\ \rm{M_{\odot}}$ (see also the alternative approach of \citealt{Kushnir2017Tides}). The calculation of $k$ and $\tau_{\rm{tide}}$ follows \cite{hurley2000comprehensive} and is uncertain for massive stars. 
For the radial expansion timescale, the fitting formulae we use are not accurate in representing the evolution of the star on thermal or dynamical timescales. These formulae also miss detailed information about the evolution of, e.g., the size of the convective envelope.
Additionally, they are not accurate in representing the effect of mass loss and mass gain. 

\begin{table*}
\caption{Properties of the donor star and the binary at the onset of \ac{RLOF} leading to a \ac{CEE}. In this Table, we list the symbols and units for each parameter, as well as the figure where the parameter is presented.}
\centering
\begin{tabular}{@{}cccc@{}}
\hline\hline
Property & Symbol & Units & Figure \\
\hline%
Luminosity & $L_{\rm{donor}}$ & $\rm{L_{\odot}}$ & \ref{fig:bigHR} \\
Effective temperature & $T_{\rm{eff,donor}}$  & K & \ref{fig:bigHR}\\
Stellar phase & - & - & \ref{fig:bigHR}  \\ 
Mass & $m_{\rm{donor}}$ & $\rm{M_{\odot}}$ & \ref{fig:donorProperties}, \ref{fig:massesAndSeparations} \\
Envelope mass & $m_{\rm{env,donor}}$ & $\rm{M_{\odot}}$ & - \\
Core mass & $m_{\rm{core,donor}}$ & $\rm{M_{\odot}}$ & - \\
Core mass fraction & $f_{\rm{donor}}\equiv m_{\rm{core,donor}}/m_{\rm{donor}}$  & - & \ref{fig:donorProperties} \\
Radial expansion timescale & $\tau_{\rm{radial,donor}}$ & $\rm{Myr}$ & \ref{fig:convectiveTidalTimescales}, \ref{fig:comparisonCircularisationTimescales}\\ 
Binding energy & $|E_{\rm{bind}}|$  & erg & \ref{fig:donorProperties},\ref{fig:Orsola} \\
Eccentricity & $e$ & - & \ref{fig:orbitalProperties} \\
Semi-major axis & $a$ & $\rm{R_{\odot}}$ & \ref{fig:Orsola},\ref{fig:massesAndSeparations} \\
Periastron & $a_{\rm{p}}= a(1-e)$ & $\rm{R_{\odot}}$ & \ref{fig:orbitalProperties},\ref{fig:massesAndSeparations} \\
Companion mass & $m_{\rm{comp}}$ & $\rm{M_{\odot}}$ & \ref{fig:massesAndSeparations} \\
Total mass & $m_{\rm{total}}= m_{\rm{donor}}+m_{\rm{comp}}$ & $\rm{M_{\odot}}$ & \ref{fig:massProperties} \\
Mass ratio & $q=m_{\rm{comp}}/m_{\rm{donor}}$ & - & \ref{fig:massProperties},\ref{fig:Orsola} \\
Circularisation timescale & $\tau_{\rm{circ}}$ & $\rm{Myr}$ & \ref{fig:convectiveTidalTimescales}, \ref{fig:comparisonCircularisationTimescales} \\
\hline\hline
\end{tabular}
\label{tab:properties}
\end{table*}

\section{Results}
\label{sec:results}
We present the results of the synthetic population of binaries which become \acp{DNS}. We focus our attention on the properties of the systems at the onset of the \ac{CEE}. If a donor star experiences \ac{RLOF}, leading to a dynamically unstable mass transfer episode, the system is classified as experiencing a \ac{CEE}. In that case we report the properties of the system at the moment of \ac{RLOF}. We do not resolve the details of the \ac{CEE}, such as the possible delayed onset of the dynamical inspiral phase. Given that we are interested in \ac{DNS} progenitors, all of these \acp{CEE} will, by selection, experience a successful ejection of the envelope, i.e. no stellar mergers are reported in this study. All the data presented in this work are available at \href{https://zenodo.org/record/3593843}{https://zenodo.org/record/3593843} \citep{VignaGomez2019CEEsDataset}.

Our synthetic data set contains about \numberOfBinaries\ binaries evolved using COMPAS.
Out of all the simulated binaries, targeted at \ac{DNS}-forming systems (see Section \ref{sec:AIS}), there are \numberOfDNSsWithCEEs\ \acp{CEE} leading to \ac{DNS} formation.  These provide a far more accurate sampling of the $\approx$\numberOfDNSsWithCEEsWeighted\  systems that would be expected for \totalMassCatalogue\ $\rm{M_{\odot}}$ of star-forming mass sampled from the initial conditions.
For simplicity, we assume 100\% binarity a priori. Nevertheless, given our assumed separation distribution that is capped at 1000 AU, \effectivelySingle\% of our systems never experience any mass transfer episode, resulting in two effectively single stars. While \acp{DNS} are believed to form in different environments, several studies have shown that metallicity does not play a large role in \ac{DNS} properties, unlike binary black hole or neutron star/black hole formation \citep{dominik2012double, VignaGomez2018DNSs,GiacobboMapelli2018,Neijssel2019}.

The results section is structured as follows. 
Section \ref{subsec:formation} discusses the two dominant formation channels in our model, i.e. the evolutionary history of the binary from \ac{ZAMS} to \ac{DNS} formation. 
Section \ref{subsec:comparison} we present a comparison with the results from \cite{VignaGomez2018DNSs}.
In Section \ref{subsec:CEEsToDNSs} we describe the way main results are presented.
In Section \ref{subsec:propertiesDonor} we report the properties of the donor. 
In section \ref{subsec:propertiesBinary} we report the properties of the binary, in particular the orbital properties.
Finally, in Section \ref{subsec:TidalTimescales} we present and report the tidal circularisation timescales.

\subsection{Formation Channels of Double Neutron Star systems} \label{subsec:formation}
Two common evolutionary pathways leading to the formation of \ac{DNS} from isolated binary evolution are identified in the literature \citep{bhattacharya1991formation,tauris2006formation,tauris2017formation}. Following \citet{VignaGomez2018DNSs}, we refer to these formation channels as \textit{Channel I} and \textit{Channel II}.\\

\begin{figure*}
\includegraphics[width=\textwidth]{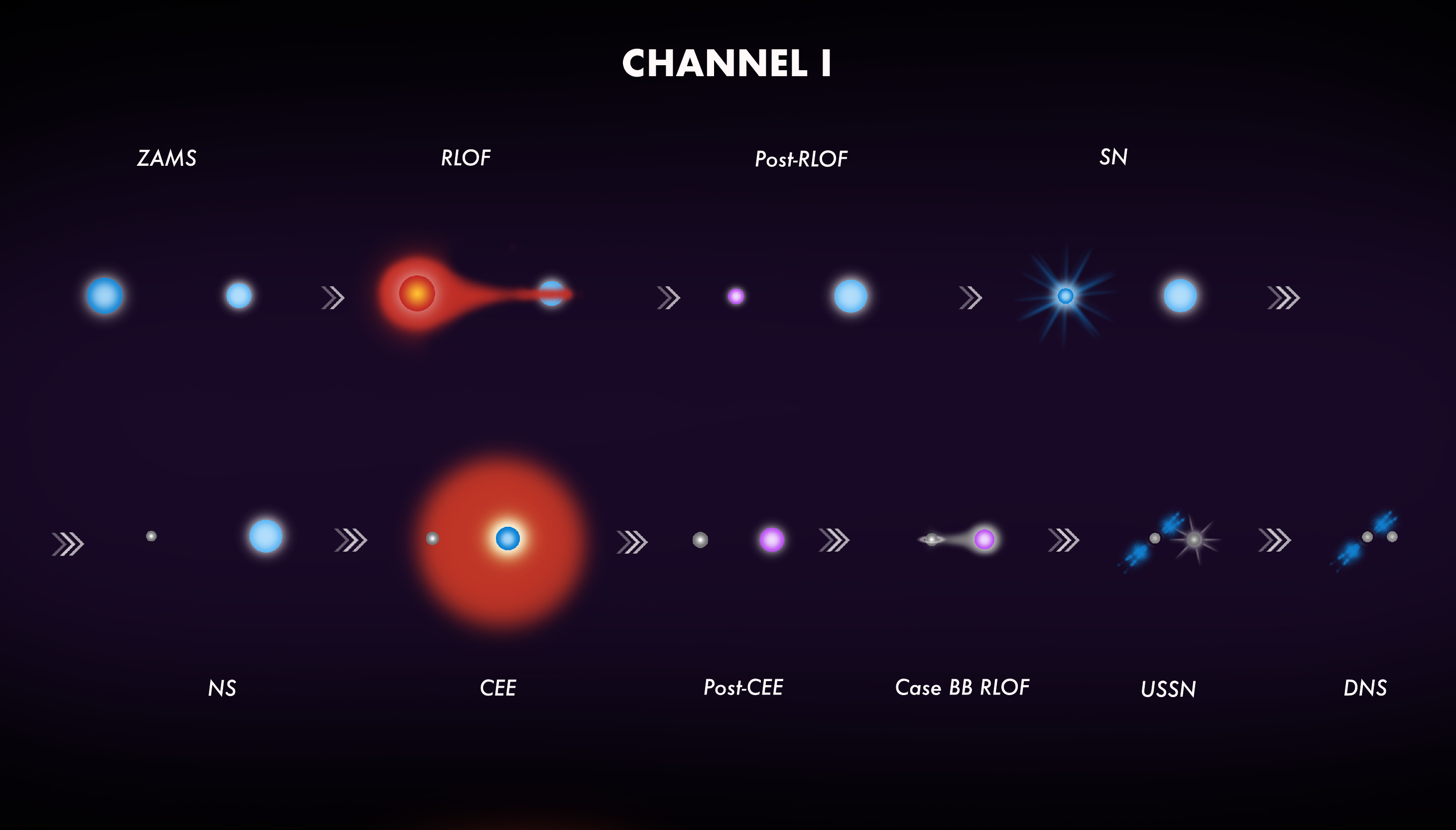}
\includegraphics[width=\textwidth]{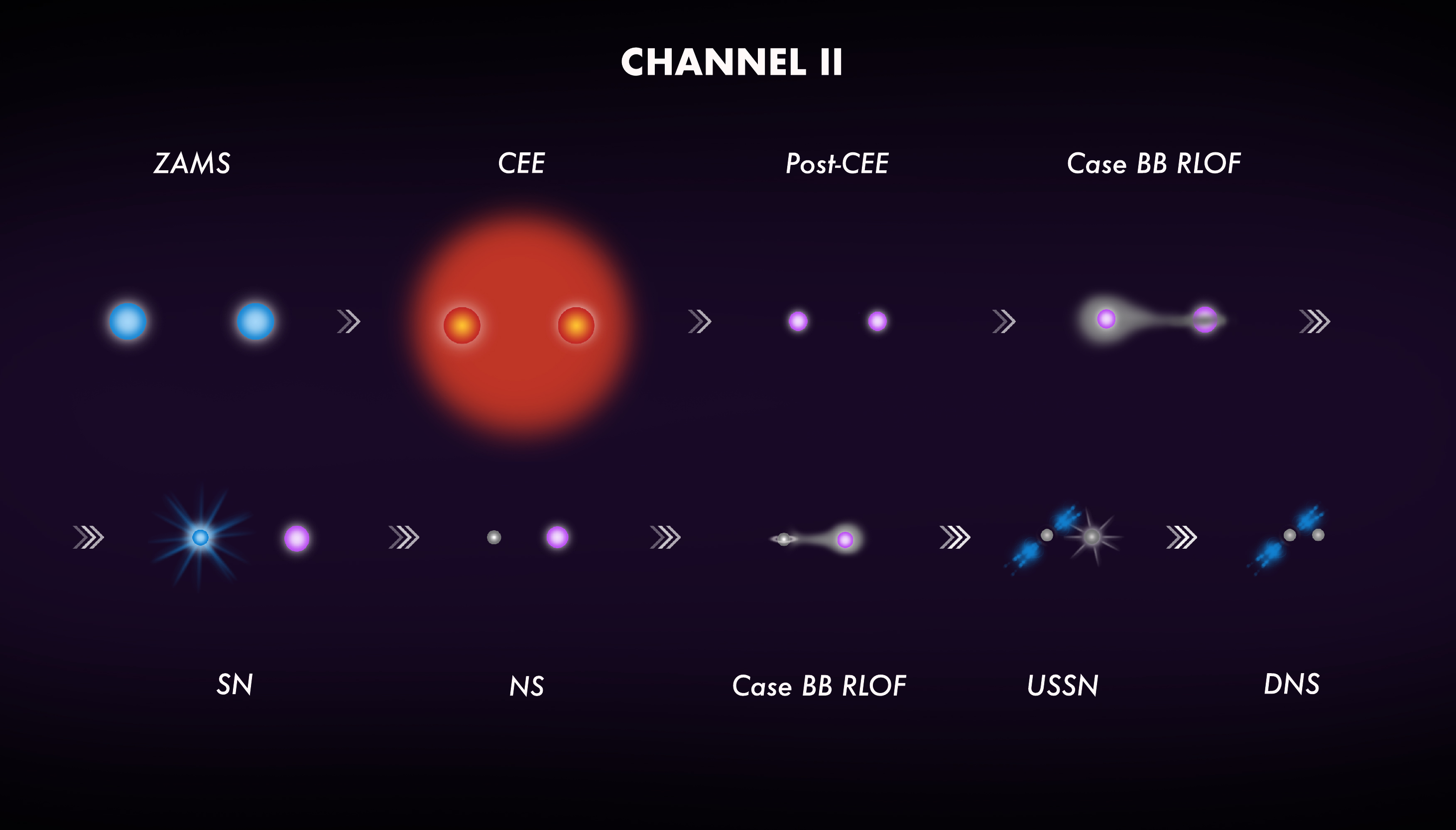}
\caption{
Schematic representation of DNS formation channels as described in Section \ref{subsec:formation}. Top: Channel I is the dominant formation channel for DNS systems, as well as the most common formation channel in the literature (see, e.g., \citealt{tauris2017formation} and references therein). Bottom: Formation Channel II distinguished by an early double-core common-envelope phase.
Acronyms as defined in text. Credit: T. Rebagliato.
}
\label{fig:channels}
\end{figure*}

Channel I is illustrated in the top panel of Figure \ref{fig:channels} and proceeds in the following way:
\begin{enumerate}
    \item A post-\ac{MS} primary engages in stable mass transfer onto a \ac{MS} secondary. 
    \item The primary, now stripped, continues its evolution as a naked helium star until it explodes in a supernova, leaving a \ac{NS} remnant in a bound orbit with a \ac{MS} companion. 
    \item The secondary evolves off the \ac{MS}, expanding and engaging in a \ac{CEE} with the \ac{NS} accretor. 
    \item After successfully ejecting the envelope, and hardening the orbit, the secondary becomes a naked helium star. 
    \item The stripped post-helium-burning secondary engages in highly non-conservative stable (case BB) mass transfer onto the \ac{NS} companion. 
    \item After being stripped of its helium envelope, the ultra-stripped secondary \citep{tauris2013ultra,tauris2015ultra} continues its evolution until it explodes as an \ac{USSN}, forming a \ac{DNS}.
\end{enumerate}

In Channel I the \ac{CEE} may occur while the donor is crossing the \ac{HG}, i.e. between the end of the \ac{MS} and the start of the \ac{CHeB} phase. Rapid population synthesis modelling of \acp{CEE}  sometimes parameterise these donors in two possible outcomes: \textit{optimistic} and \textit{pessimistic} \citep{dominik2012double}. The optimistic approach assumes the donor has a clear core/envelope separation and that, as a result, the two stellar cores can potentially remove the common envelope, allowing the binary to survive the \ac{CEE}. Throughout this paper, we assume the optimistic approach unless stated otherwise. The pessimistic approach assumes that dynamically unstable mass transfer from a \ac{HG} donor leads imminently to a merger. The pessimistic approach results in \percentageHG\% of potential \ac{DNS} candidates merging before \ac{DCO} formation.\\

Channel II is illustrated in the bottom panel of Figure \ref{fig:channels} and proceeds in the following way:
\begin{enumerate}
    \item A dynamically unstable mass transfer episode leads to a \ac{CEE} when the primary and the secondary are both post-\ac{MS} star. 
    During this \ac{CEE}, both stars have a clear core-envelope separation, and they engage in what is referred to in the literature as a double-core \ac{CEE} \citep{brown1995doubleCore,dewi2006double,Justham+2011}.
    For these binaries, evolutionary timescales are quite similar, with a minimum and mean mass ratio of $\approx0.93$ and $\approx 0.97$ respectively, consistent with high-mass and low-mass solar metallicity values reported in \cite{dewi2006double}.
    During this double-core \ac{CEE}, both stars are stripped and become naked-helium-stars. 
    \item The stripped post-helium-burning primary engages in stable (case BB) mass transfer onto a stripped helium-burning secondary. 
    \item The primary, now a naked metal star, explodes in a \ac{SN} and becomes a \ac{NS}. 
    \item There is a final highly non-conservative stable (case BB) mass transfer episode from the stripped post-helium-burning secondary onto the \ac{NS}.
    \item The secondary then explodes as an \ac{USSN}, forming a \ac{DNS}.
\end{enumerate}

The two dominant channels, Channel I and Channel II, comprise \fractionChannelI\ and \fractionChannelII\ of all \acp{DNS} in our simulations ($Z=0.0142$), respectively.
The remaining formation channels are mostly variations of the dominant channels. These variations either alter the sequence of events or avoid certain mass transfer phases. Some formation scenarios rely on fortuitous \ac{SN} kicks. Some other exotic scenarios, which allow for the formation of \ac{DNS} in which neither neutron star is recycled by accretion \citep[e.g. ][]{1538-4357-550-2-L183}, comprise less than 2\% of the \ac{DNS} population.

\begin{figure*}
\includegraphics[width=\textwidth]{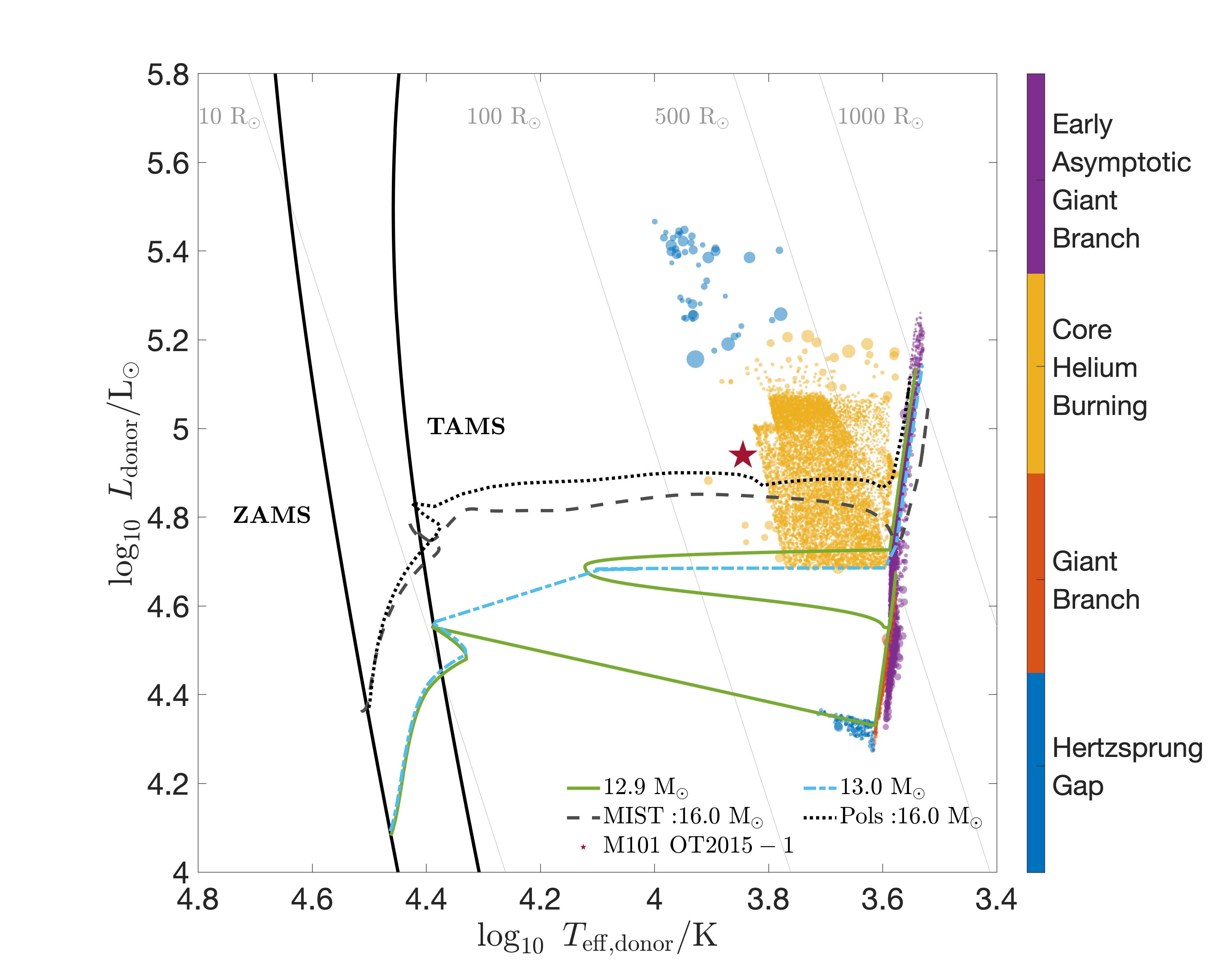}
\includegraphics[trim={1cm 0 1cm 0},clip,width=0.329\textwidth]{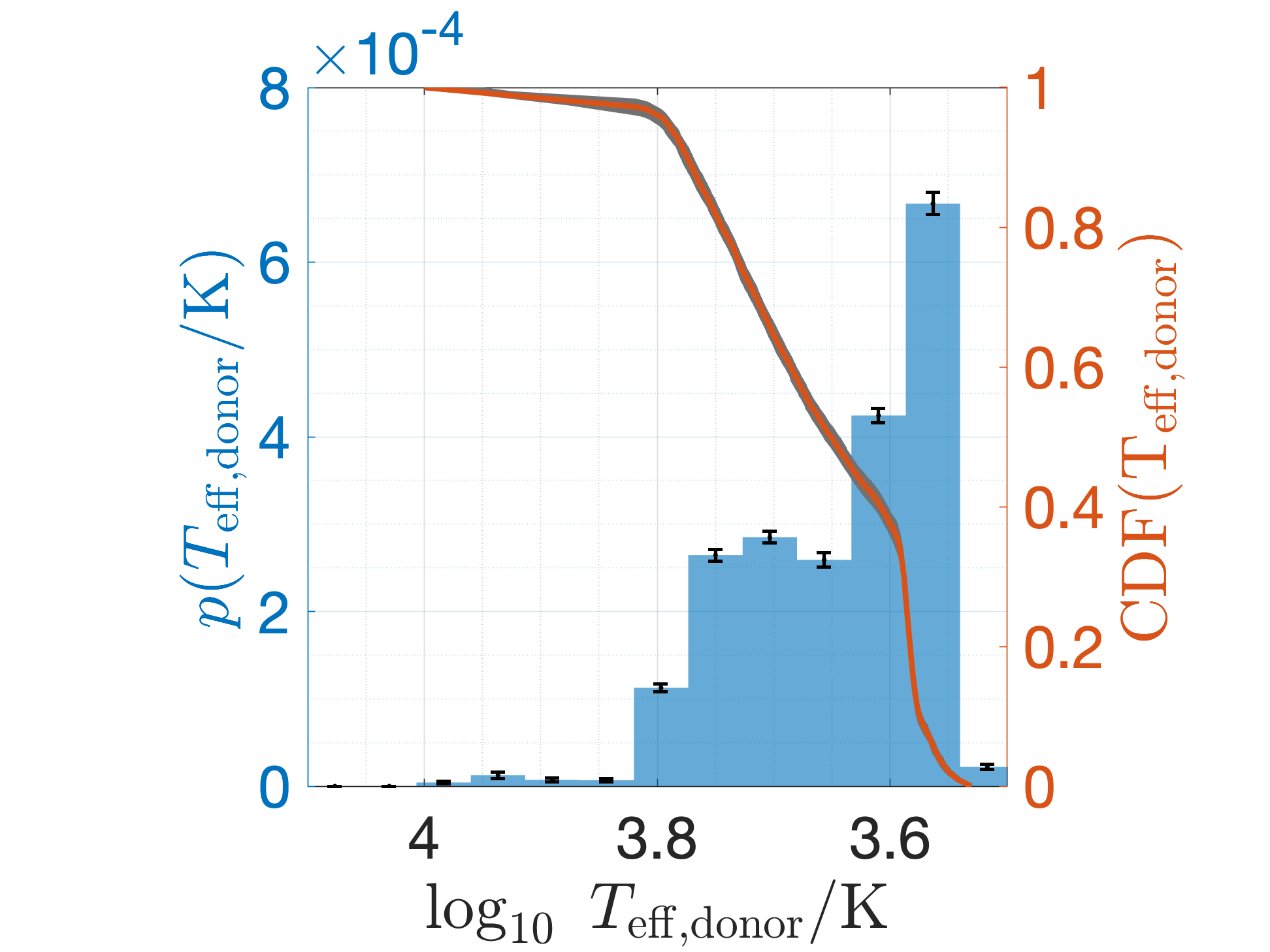}
\includegraphics[trim={1cm 0 1cm 0},clip,width=0.329\textwidth]{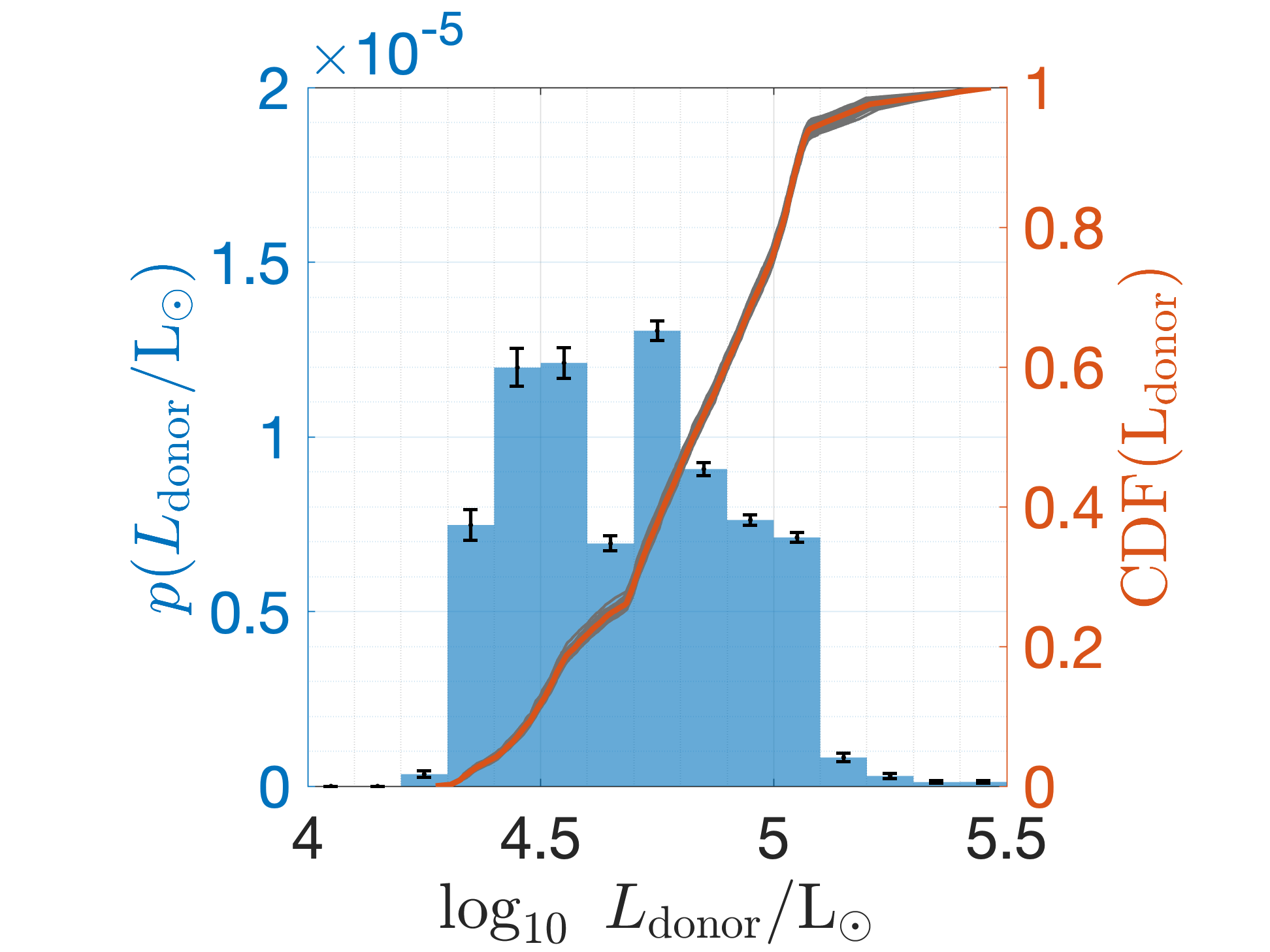}
\includegraphics[trim={1cm 0 1cm 0},clip,width=0.329\textwidth]{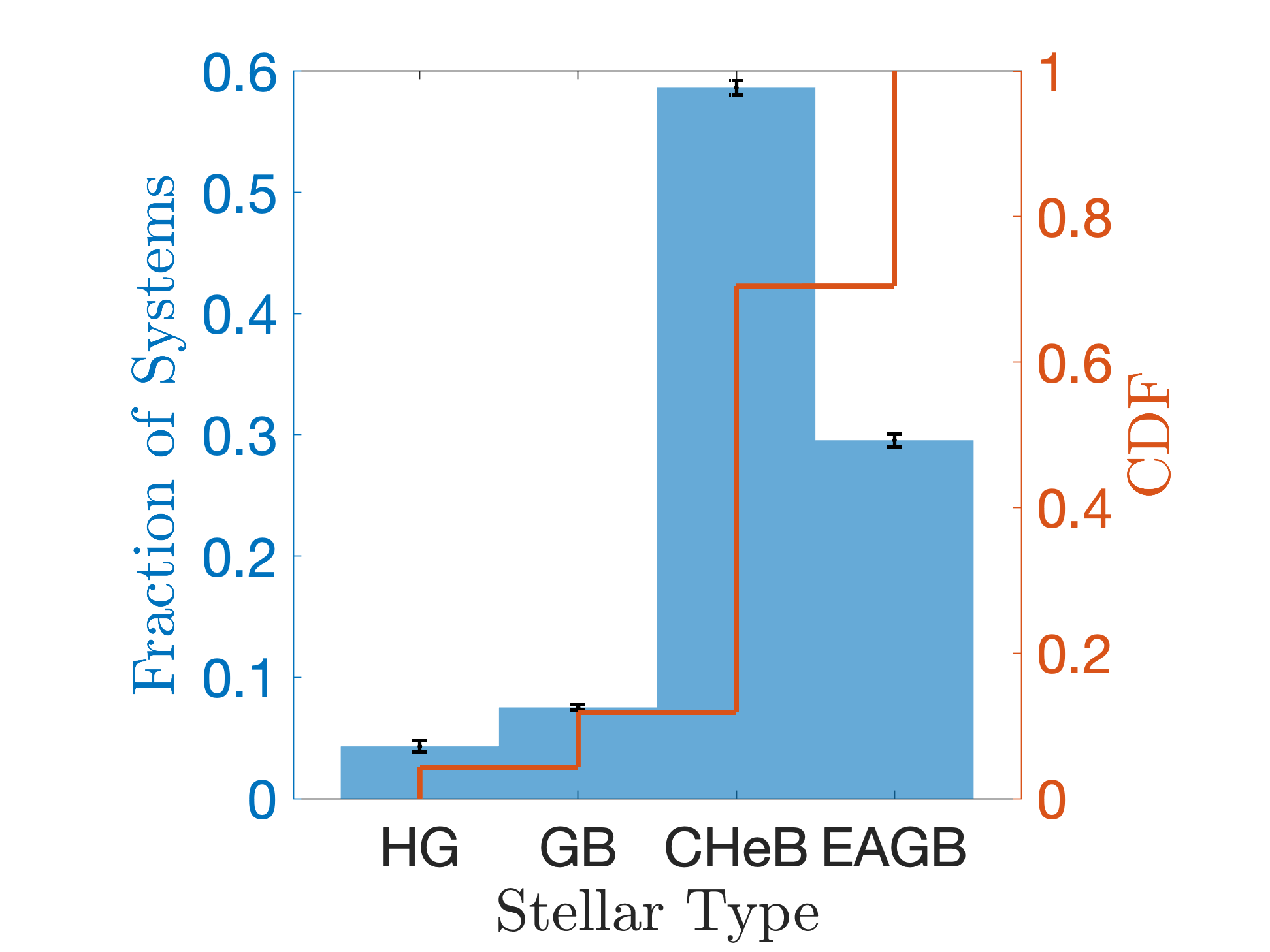}
\caption[HR diagram of CEEs leading to DNSs]{
Main properties of the donor star at the onset of \ac{RLOF} leading to the \ac{CEE} in \ac{DNS}-forming binaries. 
Top: \ac{HR} diagram coloured by stellar phase: \ac{HG} (blue), \ac{GB} (orange), \ac{CHeB} (yellow) and EAGB (purple). The sizes of the markers represent their sampling weight. We show the progenitor of the luminous red nova M101 OT2015-1 \citep{Blagorodnova2017LRN} with a star symbol. 
The solid black lines indicate \ac{ZAMS} and \ac{TAMS} loci for a grid of SSE models \citep{hurley2000comprehensive} at $Z\approx0.0142$. We show the evolution of a single non-rotating $16\ \rm{M_{\odot}}$ star, from \ac{ZAMS} to the end of the giant phase: the dotted dark-grey line shows a MIST stellar track from \cite{Choi+2016MIST} and the dashed grey line shows the stellar track from \cite{pols1998stellar,Pols2009}. The dash-dotted light-blue and solid green lines show how fitting formulae from \cite{hurley2000comprehensive} lead to a bifurcation after the \ac{MS} for stars with masses between 12.9 and 13.0 $M_\odot$. 
This bifurcation is related to which stars are assumed to begin core-helium-burning while crossing of the \ac{HG} or only after it: see the presence (lack) of the blue loop in the $12.9\ (13.0)\ \rm{}M_{\odot}$ track.
Grey lines indicate stellar radii of $R=\{10,100,500,1000\}\ \rm{R_{\odot}}$.
Bottom: Normalised distributions in blue (left vertical axis) and \acp{CDF} in orange (right vertical axis) of luminosity (left panel), effective temperature (middle panel) and stellar type (right panel). Black error bars indicate $1\sigma$ sampling uncertainty in the histograms. 
Grey lines show 100 bootstrapped distributions that indicate the sampling uncertainty in the CDFs.
The CDFs show a subset of 365 randomly sampled values, which is the same number of \ac{DNS} in our population, for each bootstrapped distribution.
}
\label{fig:bigHR}
\end{figure*}

\subsection{Comparison with \cite{VignaGomez2018DNSs}}\label{subsec:comparison}
This work generally uses similar assumptions and physics parameterisations as the preferred model of \citet{VignaGomez2018DNSs}, including the \citet{fryer2012compact} delayed supernova engine.  Although the qualitative results are similar, there are some quantitative changes due to updated model choices and corrections to the COMPAS population synthesis code as described in Section \ref{subsec:underlyingPhysics} and Appendix \ref{app:popSynth}. For example, the percentage of systems forming though Channel I remains $\approx 70 \%$, but now only $\approx 14\%$ of systems experience Channel II,  instead of $\approx 21 \%$ in \citet{VignaGomez2018DNSs}.  The main change concerns the \ac{DNS} rates which in this work are a factor of a few lower than those in the preferred model of \cite{VignaGomez2018DNSs}. 
Here, we estimate the formation rate of all (merging) \ac{DNS} to be \AllBNSRateGpcYear(\MergingBNSRateGpcYear) $\rm Gpc^{-3}\ yr^{-1}$. \cite{VignaGomez2018DNSs} reports the formation rate of all (merging) \ac{DNS} to be 369(281) $\rm Gpc^{-3}\ yr^{-1}$ for the preferred model.  We discuss rates in more detail in Section \ref{subsubsec:rates}.

\subsection{Common-Envelope Episodes leading to Double Neutron Star Formation in the Hertzsprung-Russell Diagram}\label{subsec:CEEsToDNSs}
For all properties, we present a colour coded \ac{HR} diagram, normalised distribution and \ac{CDF}.
In Figure \ref{fig:bigHR}, we present our synthetic population of \acs{DNS} progenitors at the onset of \ac{RLOF} leading to a \ac{CEE}. They are coloured according to the stellar type of the donor at \ac{RLOF}, which is specified using the nomenclature from \citet{hurley2000comprehensive}\footnote{We use the \ac{EAGB} nomenclature even for stars with masses $m \gtrapprox 10\ \rm{M_{\odot}}$ which do not become AGB stars.}. Additionally, Figure \ref{fig:bigHR} shows the normalised distributions of luminosity ($L_{\rm{donor}}$), effective temperature ($T_{\rm{eff,donor}}$) and stellar type of the donor. In the case of a double-core \ac{CEE}, the donor is defined as the more evolved star from the binary, which is the primary in Channel II.

In Figure \ref{fig:bigHR}, there is a visually striking feature: the almost complete absence of systems which forms a white polygon around $\log_{10}(L_{\rm{donor}}/\rm{L_{\odot}})=4.5$. This feature is a consequence of the fitting formulae used for single stellar evolution (c.f. Figures 14 and 15 of \citealt{hurley2000comprehensive}). This white region is bounded by the evolution of a $12.9\ \rm{M_{\odot}}$ and a $13.0\ \rm{M_{\odot}}$ star at Z=0.0142. The \ac{MS} evolution of both stars is quite similar. After the end of the \ac{MS}, there is a bifurcation point arising from the lower mass system experiencing a blue loop and the higher mass system avoiding it. This bifurcation is enhanced by the sharp change in the $T_{\rm{eff}}-L$ slope from the interpolation adopted by \cite{hurley2000comprehensive} during the \ac{HG} phase. This change in slope around $\log_{10}(L_{\rm{donor}}/\rm{L_{\odot}})=4.5$ and $\log_{10}(T_{\rm{eff,donor}}/\rm{K})=4.4$ is model-dependent, but we do expect to have some differences in the evolution of stars around that mass. This bifurcation corresponds to the transition around the First Giant Branch, which separates intermediate-mass and high-mass stars. Stellar tracks from \cite{Choi+2016MIST} also display a bifurcation point, but models and interpolation are smoother than those in \cite{hurley2000comprehensive}.

A rare example of how a system could end up in the forbidden region is the following. If a star experiences a blue loop, it contracts and then re-expands before continuing to evolve along the giant branch. If the companion experiences a supernova with a suitable kick while the star is in this phase, and the orbit is modified appropriately in the process, the system may experience \ac{RLOF}. A fortuitous kick making the orbit smaller, more eccentric, or both, would be an unusual but not implausible outcome of a \ac{SN}.

\subsection{Properties of the Donor}\label{subsec:propertiesDonor}
We report the luminosity, effective temperature, stellar phase, mass and core mass fraction of the donor ($f_{\rm{donor}}\equiv m_{\rm{core,donor}}/m_{\rm{donor}}$), as presented in Table \ref{tab:properties}. The luminosity and effective temperature limits are $\log_{10}\ [L_{\rm{donor,min}}/\rm{L_{\odot}},L_{\rm{donor,max}}/\rm{L_{\odot}}] = [ \lowLimLuminosity,\uppLimLuminosity]$ and $\log_{10}\ [T_{\rm{eff,donor,min}}/\rm{K},T_{\rm{eff,donor,max}}/\rm{K}] = [\lowLimTemperature,\uppLimTemperature]$, respectively. In Figure \ref{fig:bigHR} we highlight the stellar phase, which is colour-coded. While the evolution in the \ac{HR} diagram is itself an indicator of the evolutionary phase of the star, our stellar models follow closely the stellar-type nomenclature as defined in \citet{hurley2000comprehensive}. Donors which engage in a \ac{CEE} leading to \ac{DNS} formation can be in the \ac{HG} (\percentageHG\%), \ac{GB} (\percentageGB\%), \ac{CHeB} (\percentageCHeB\%) or \ac{EAGB} (\percentageEAGB\%) phase.

In the case of Channel I, donors are \ac{HG} or \ac{CHeB} stars; they span most of the parameter space from terminal-age \ac{MS} until the end of core-helium burning, with a temperature range of $\sim 0.5$ dex. In the case of Channel II, donors are \ac{GB} or \ac{EAGB} giant-like stars. The parameter space in the \ac{HR} diagram for these giant-like donors is significantly smaller, spanning an effective temperature range of only $\sim 0.1$ dex.

\begin{figure*}
\includegraphics[width=0.5\textwidth]{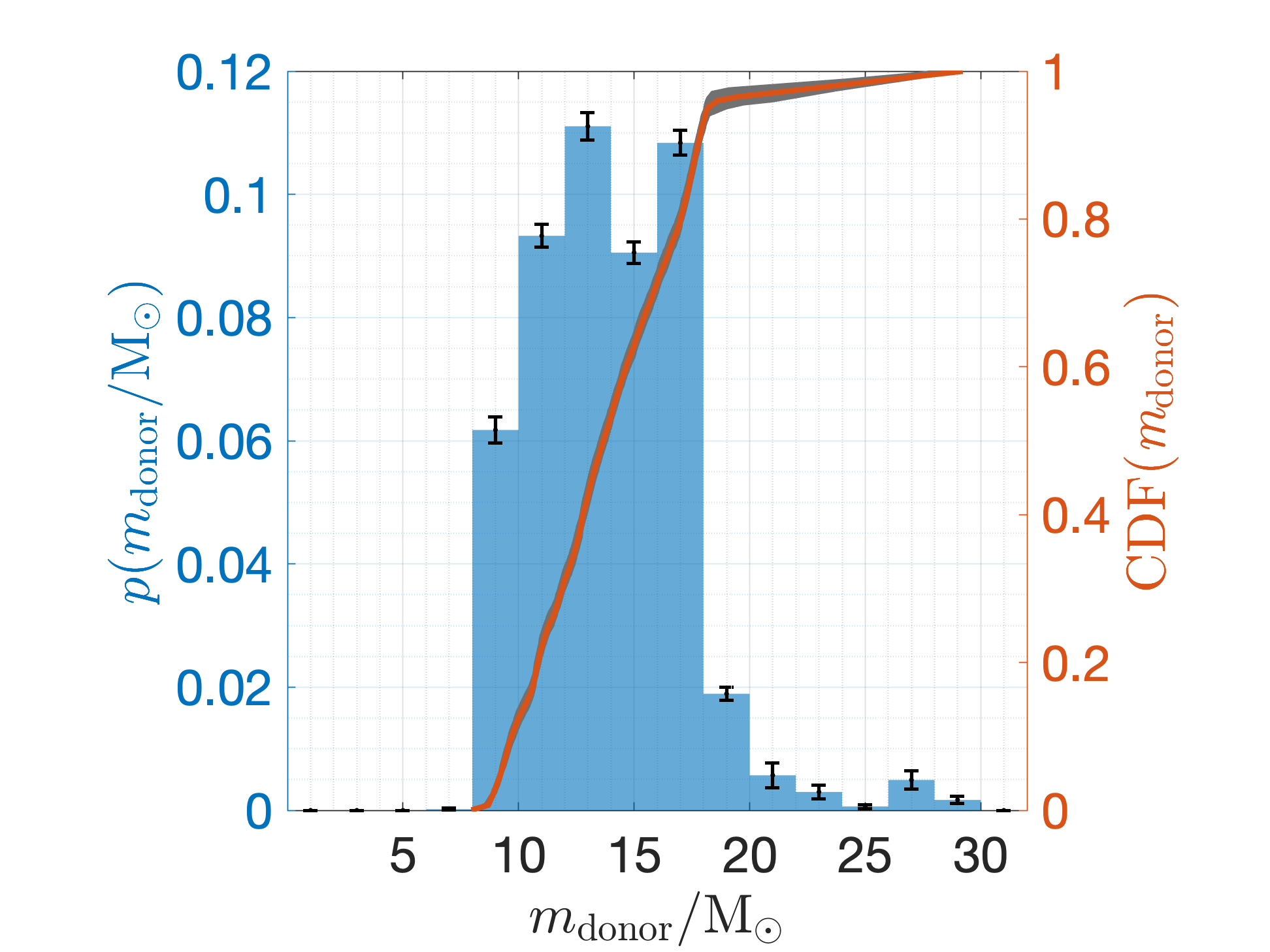}
\includegraphics[width=0.5\textwidth]{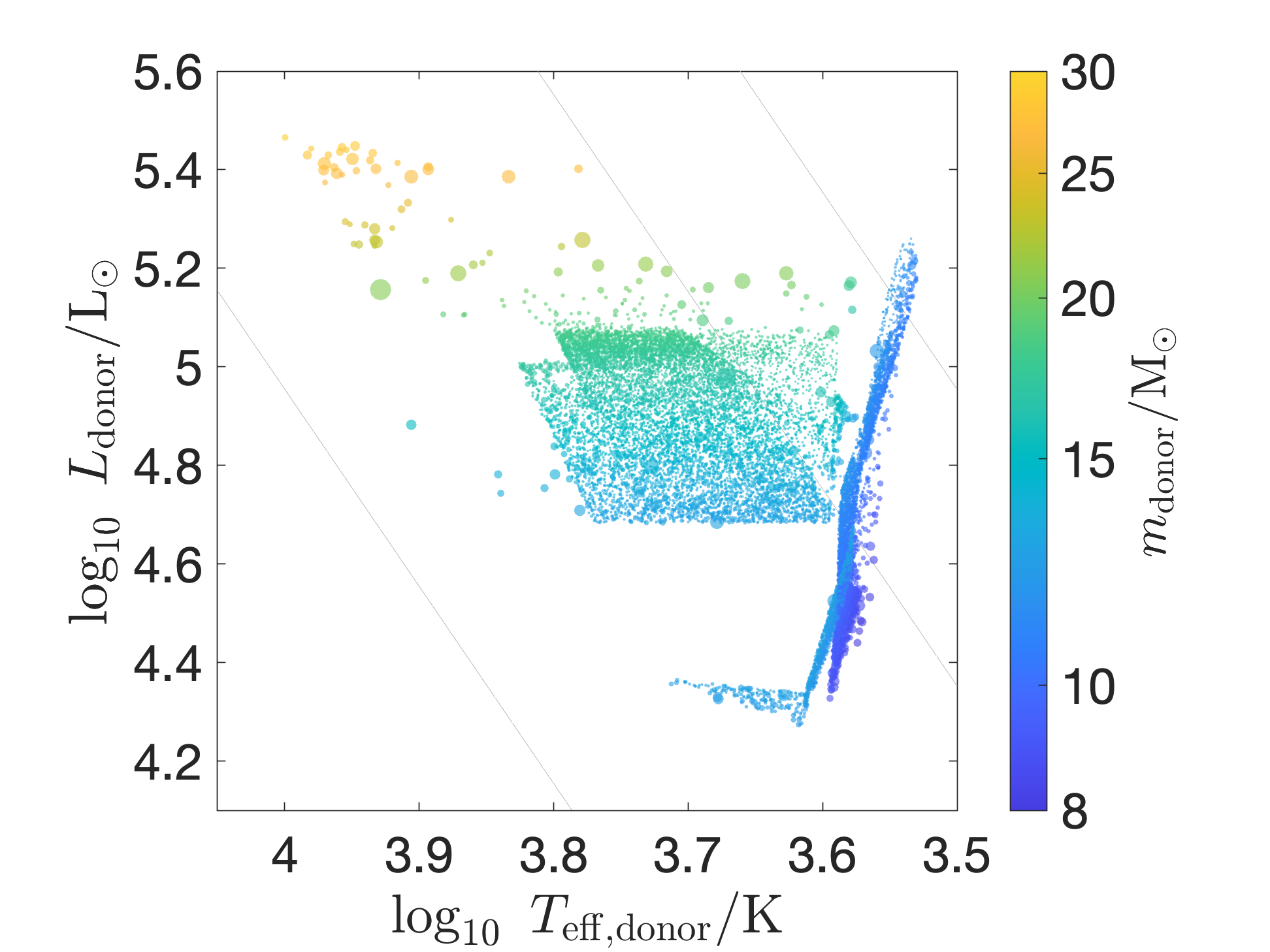}
\includegraphics[width=0.5\textwidth]{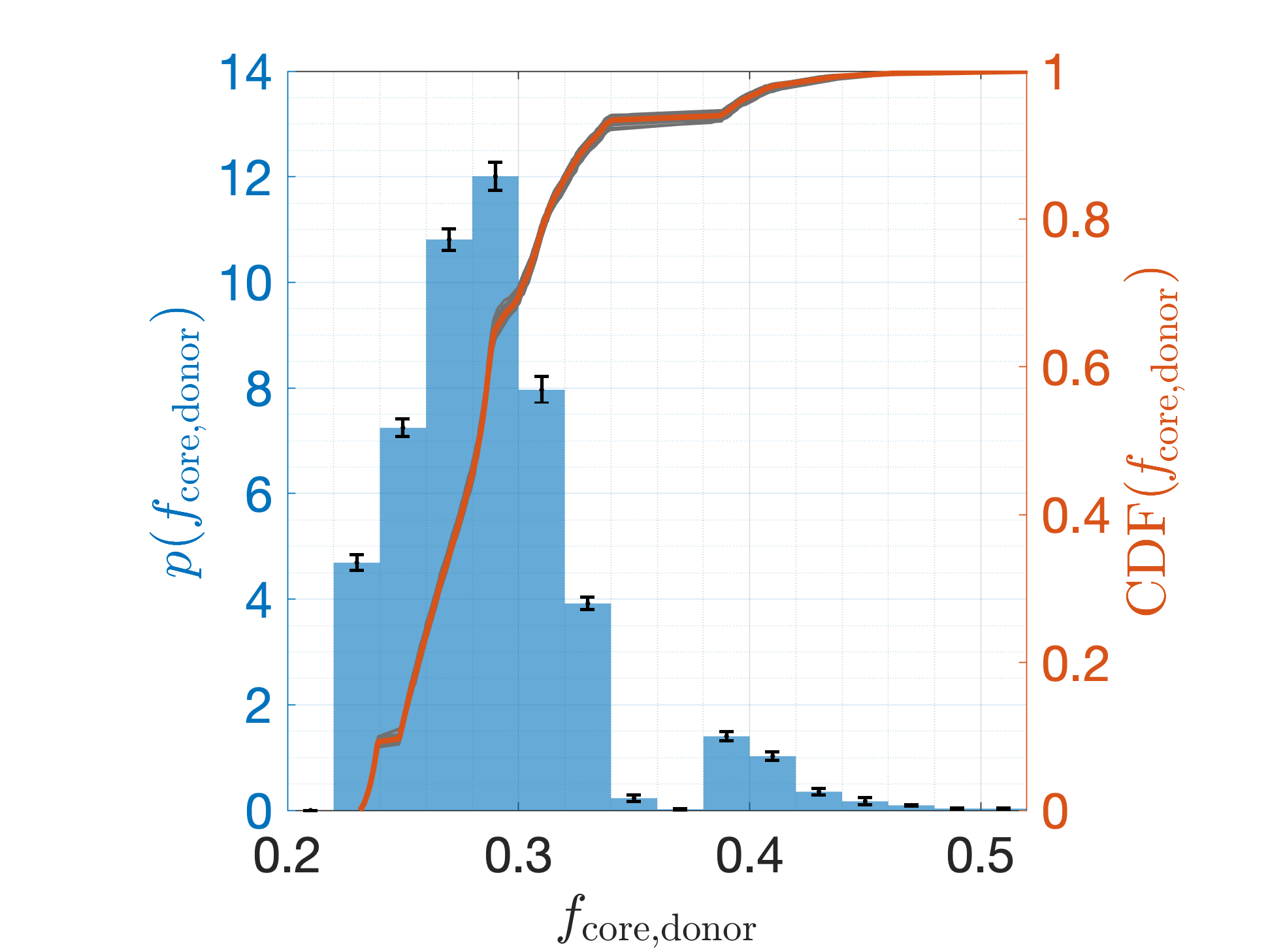}
\includegraphics[width=0.5\textwidth]{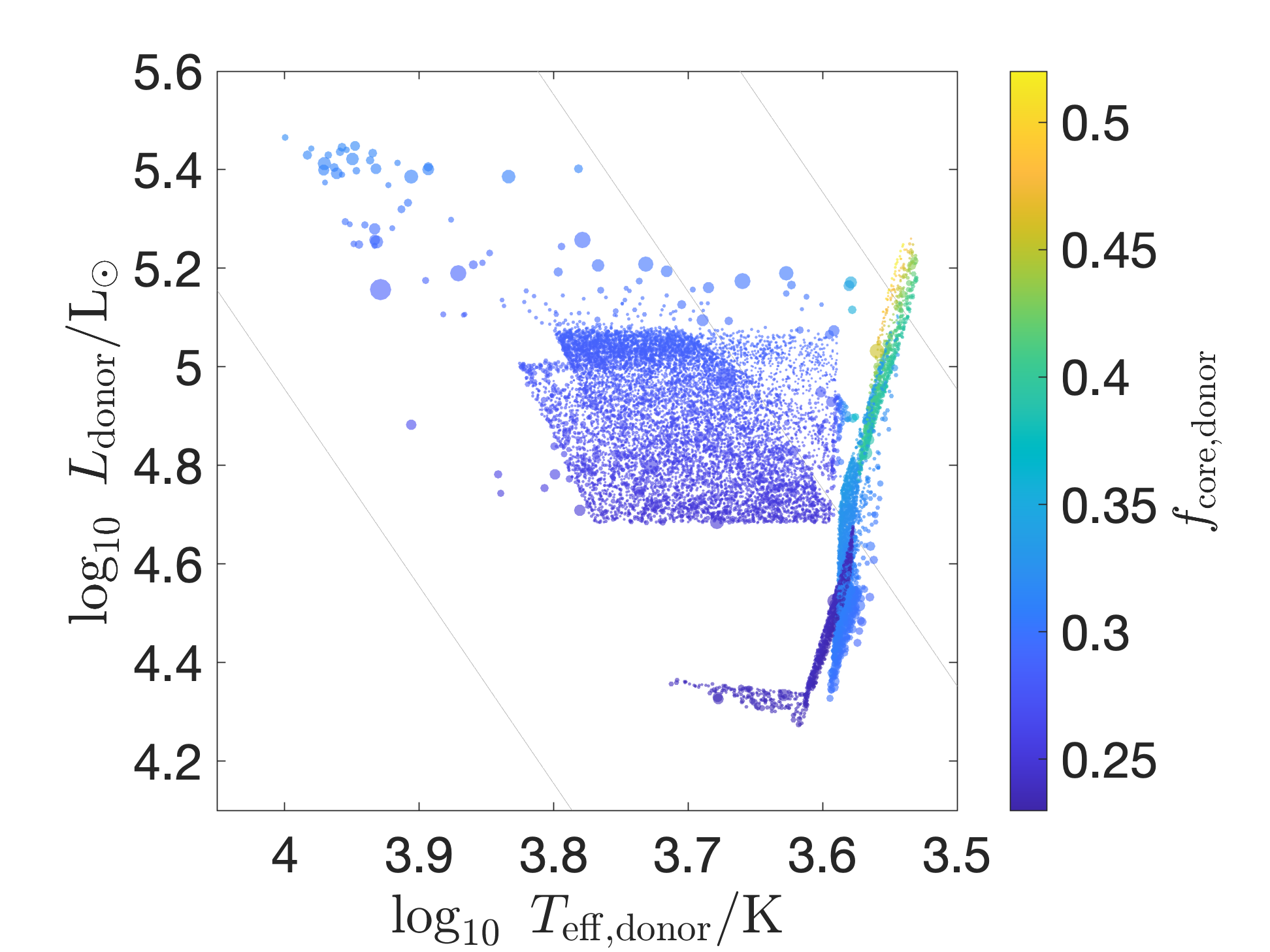}
\includegraphics[width=0.5\textwidth]{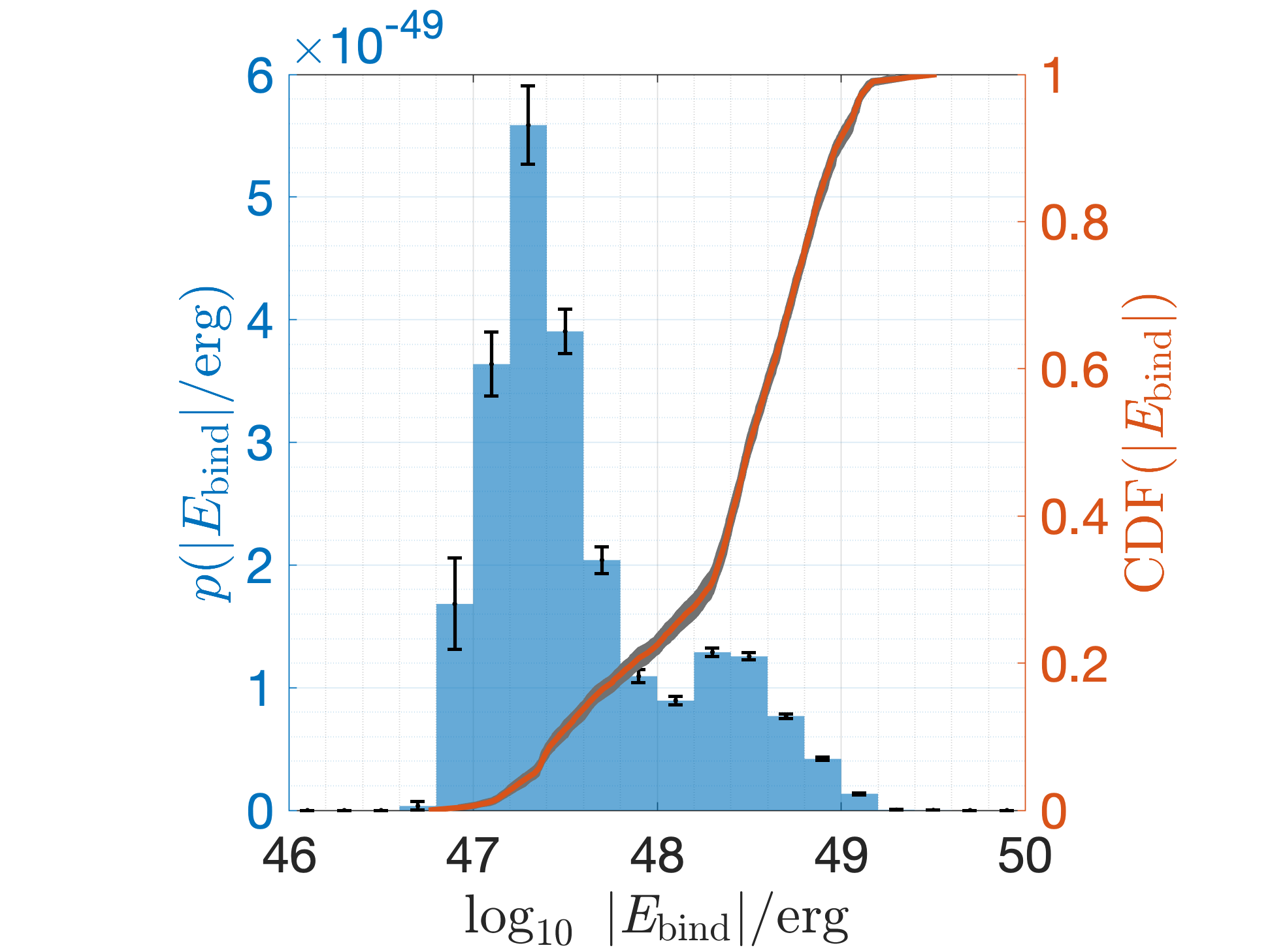}
\includegraphics[width=0.5\textwidth]{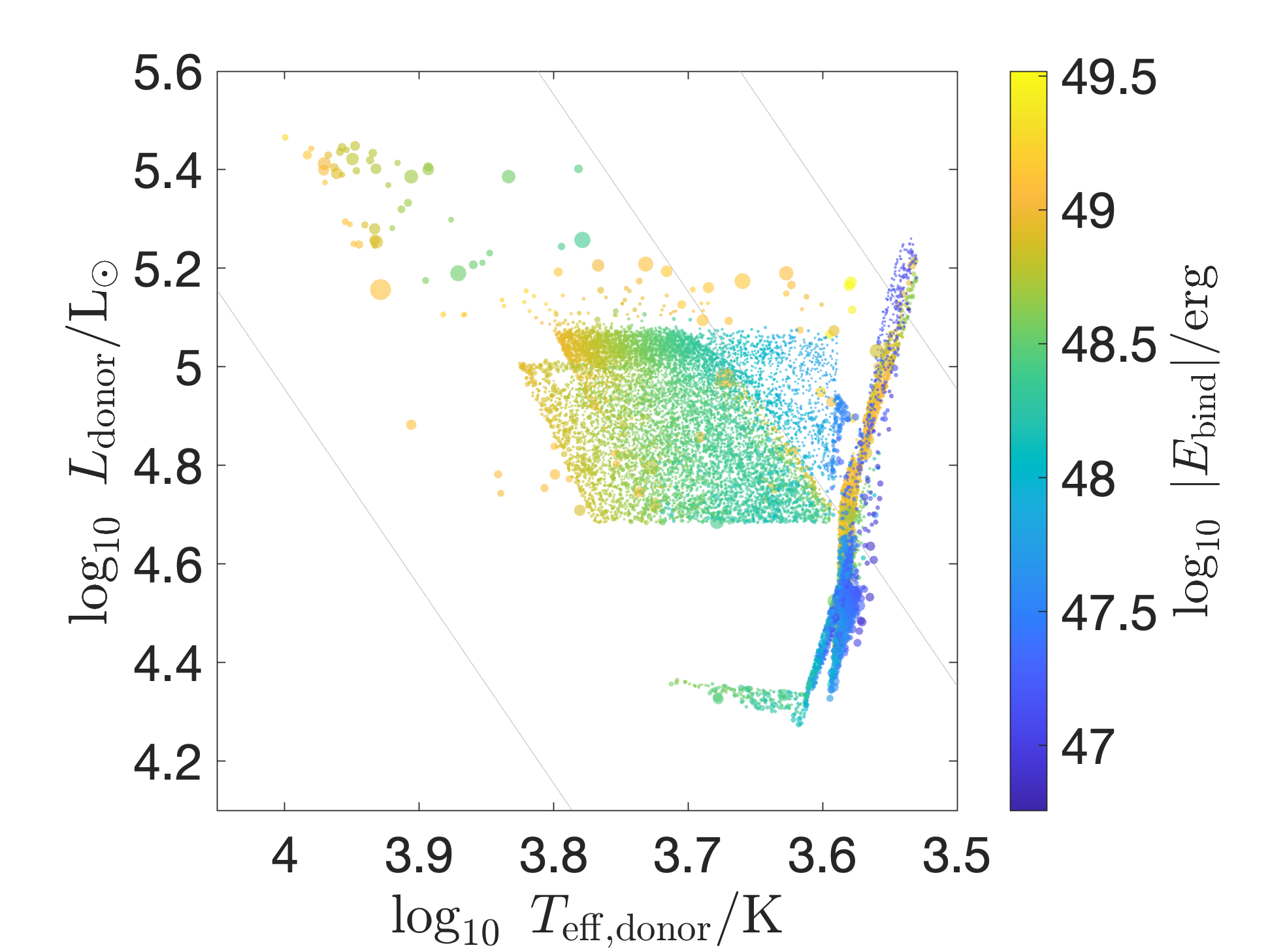}
\caption[Pre-CEE donor properties for all DNS-forming systems]{
Pre-CEE donor properties of all DNS-forming systems: mass (top), core mass fraction (middle), and envelope binding energy (bottom). The core mass fraction is defined as $f_{\rm{core,donor}}\equiv m_{\rm{core,donor}}/m_{\rm{donor}}$. 
In the case of a double-core \ac{CEE}, the binding energy is the sum of the individual envelope binding energies. Yellow systems with binding energies larger than $\log_{10}\ |E_{\rm{bind}}/\mathrm{erg}|\approx 48.5$ during the red supergiant phase are double-core \ac{CEE} systems.
For more details, see Section \ref{subsec:propertiesDonor}.
See the caption of Figure \ref{fig:bigHR} for further explanations.
}
\label{fig:donorProperties}
\end{figure*}

The limits in the mass of the donor are $[m_{\rm{donor,min}},$ $ m_{\rm{donor,max}}] =[\lowLimMass, \uppLimMass]\ \rm{M_{\odot}}$. The core mass fraction, shown in Figure \ref{fig:donorProperties}, has limits of $[f_{\rm{core,donor,min}},$ $ f_{\rm{core,donor,max}}]=[\lowLimCoreMassFraction, \uppLimCoreMassFraction]$. The core mass fraction can serve as a proxy for the evolutionary phase.

We report the binding energy of the envelope (see Figure \ref{fig:donorProperties}) as defined in Equation \ref{eq:eBind}. In the case of a double-core \ac{CEE}, the binding energy of the common envelope is assumed to be $E_{\rm{bind}}=E_{\rm{bind,donor}}+E_{\rm{bind,comp}}$. The binding energy falls in the range $\log_{10}\ (-[E_{\rm{bind,min}},E_{\rm{bind,max}}]/\rm{erg}) =[\uppLimEbind,\lowLimEbind]$. The systems with the most tightly bound envelopes, and therefore the lowest (most negative) binding energies, are those experiencing a \ac{CEE} shortly after \ac{TAMS} or as double-core \acp{CEE}. For double-core systems, the envelope of the less evolved companion is more bound than the one of the more evolved donor star. 

\begin{figure*}
\includegraphics[width=0.5\textwidth]{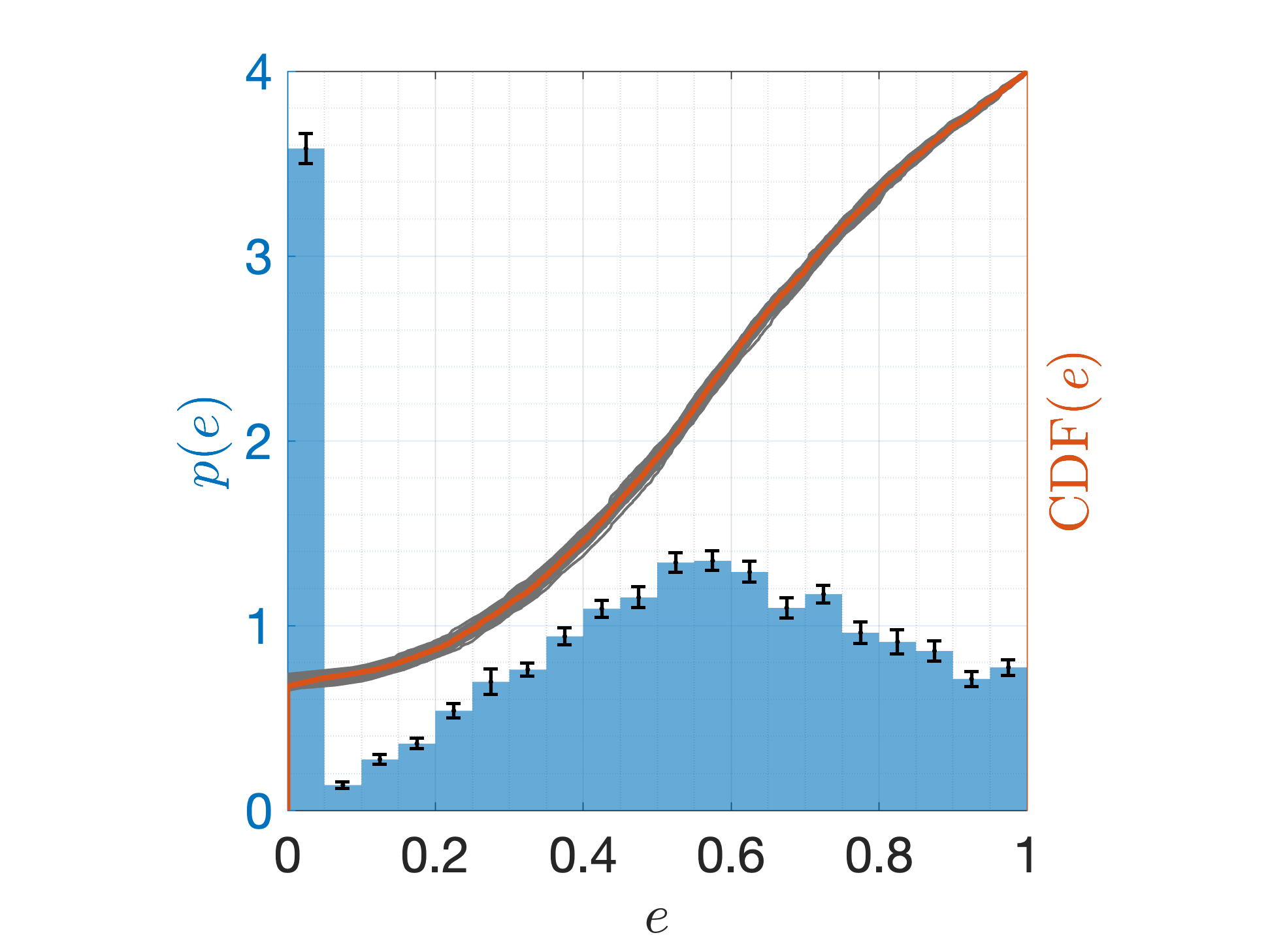}
\includegraphics[width=0.5\textwidth]{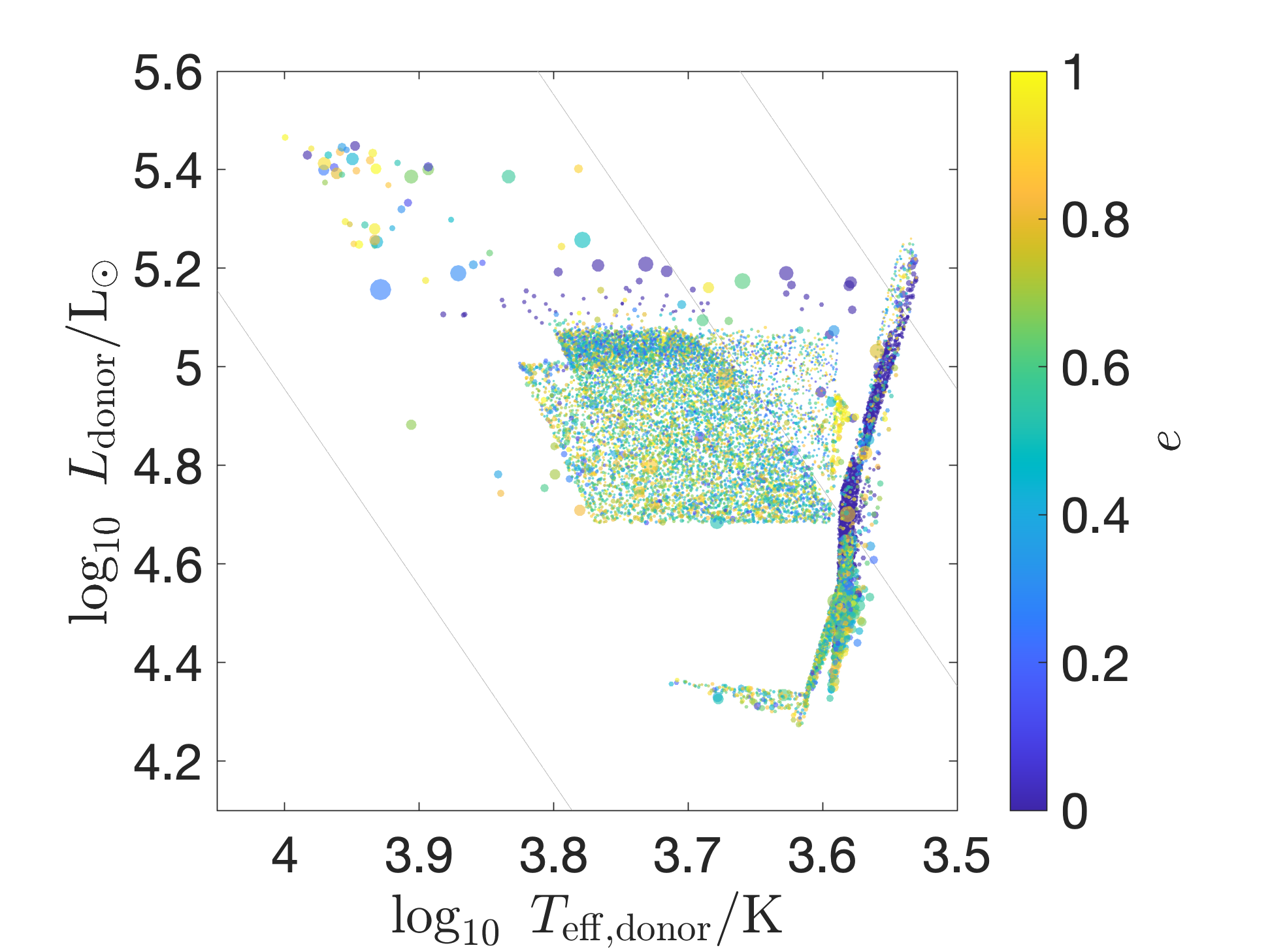}
\includegraphics[width=0.5\textwidth]{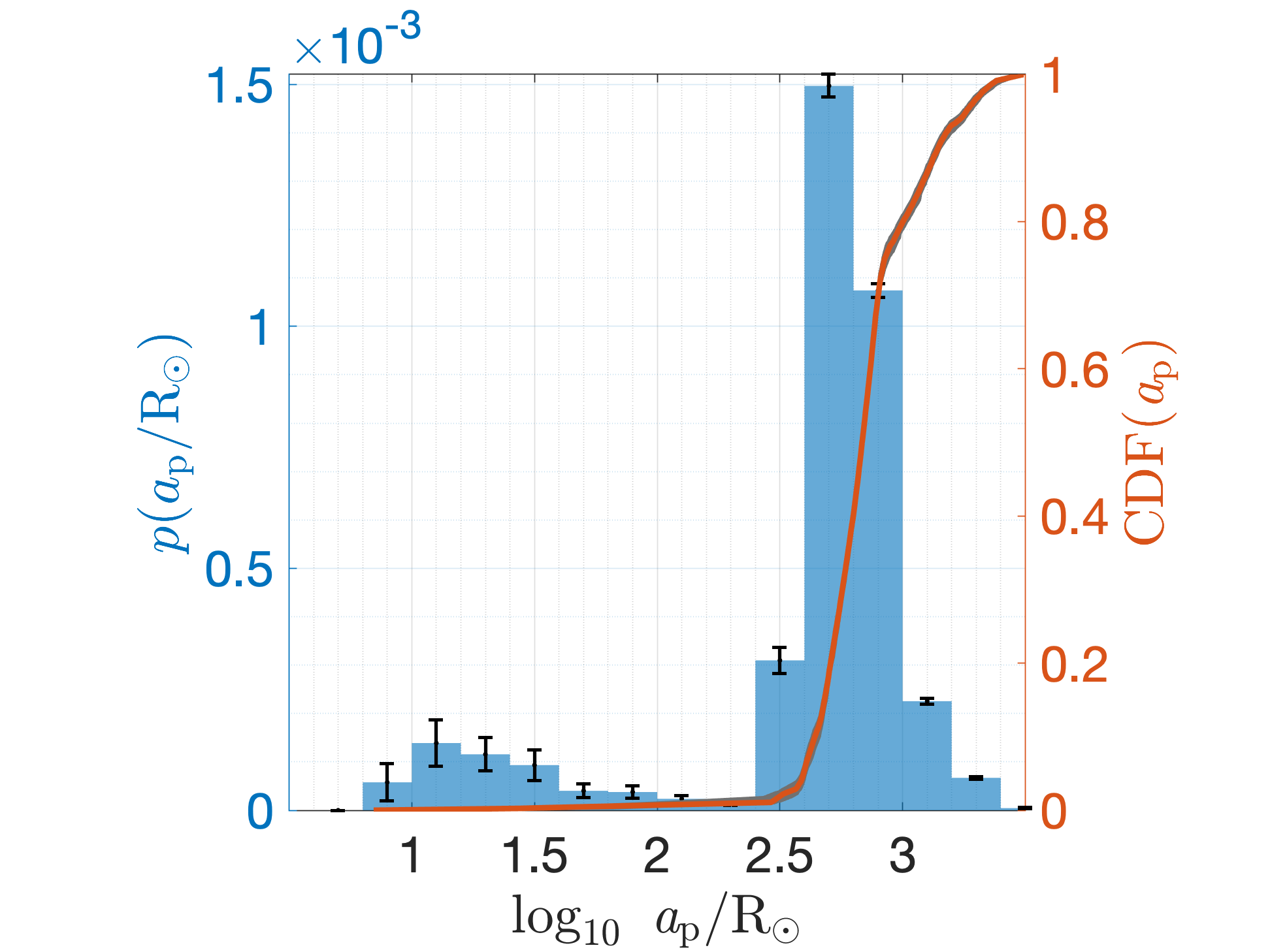}
\includegraphics[width=0.5\textwidth]{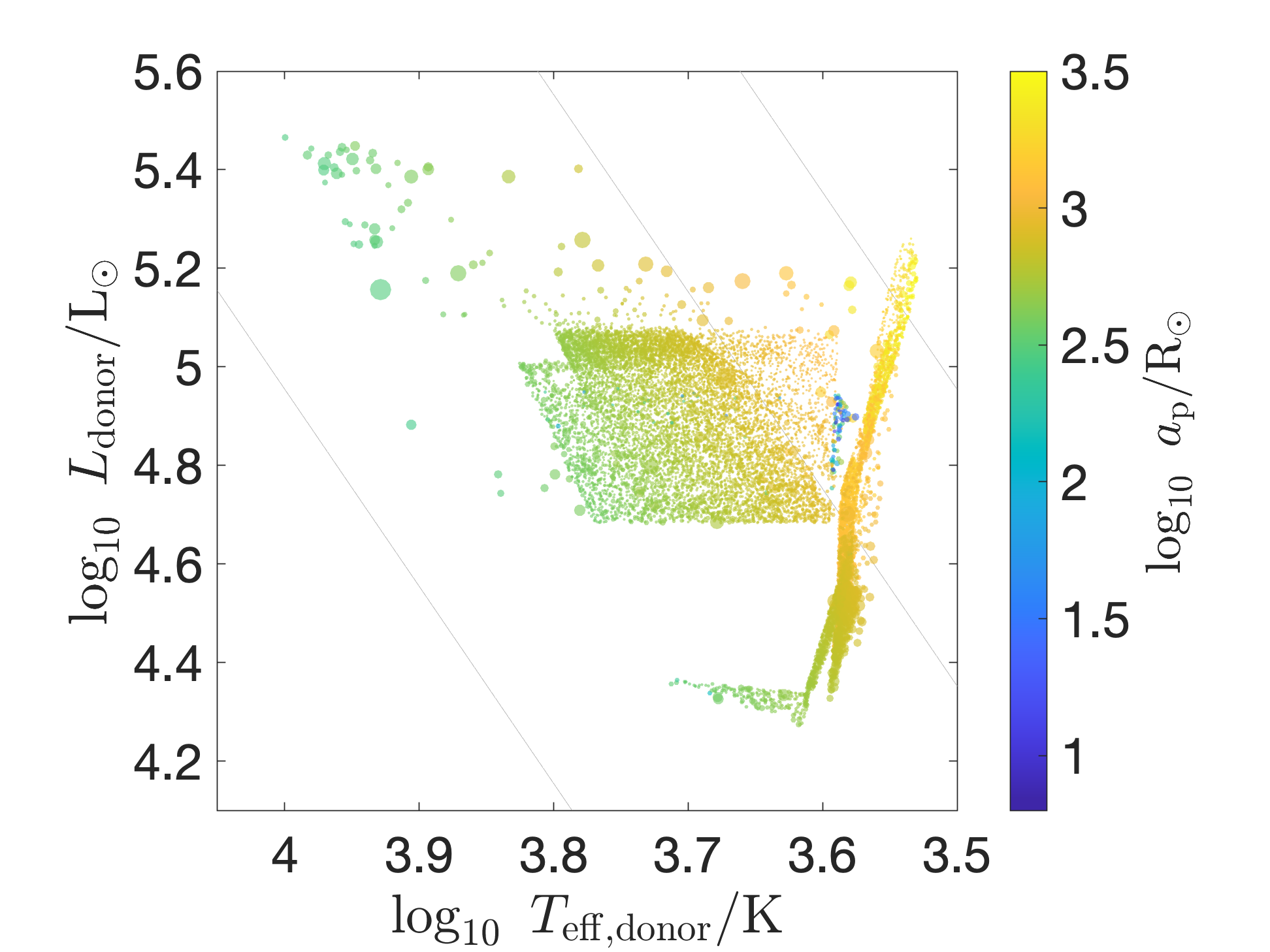}
\caption[Pre-CEE orbital properties for all DNS-forming systems]{
Pre-CEE orbital properties of all DNS-forming systems. The binary properties presented are eccentricity (top) and semi-major axis (bottom). The orbital properties do not account for tidal circularisation.
For more details, see Section \ref{subsec:propertiesBinary}.
See the caption of Figure \ref{fig:bigHR} for further explanations.
}
\label{fig:orbitalProperties}
\end{figure*}	

\begin{figure*}
\includegraphics[width=0.5\textwidth]{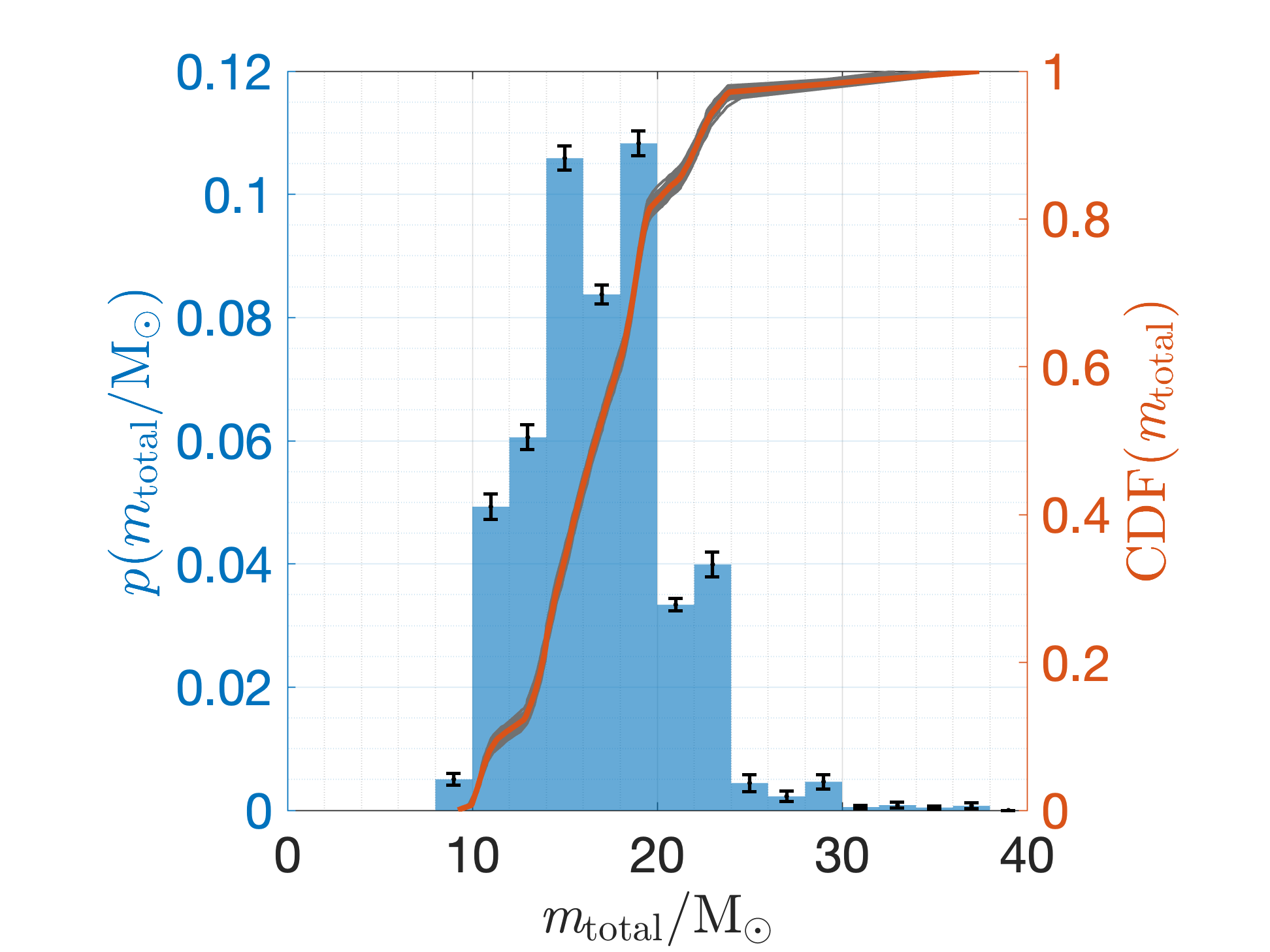}
\includegraphics[width=0.5\textwidth]{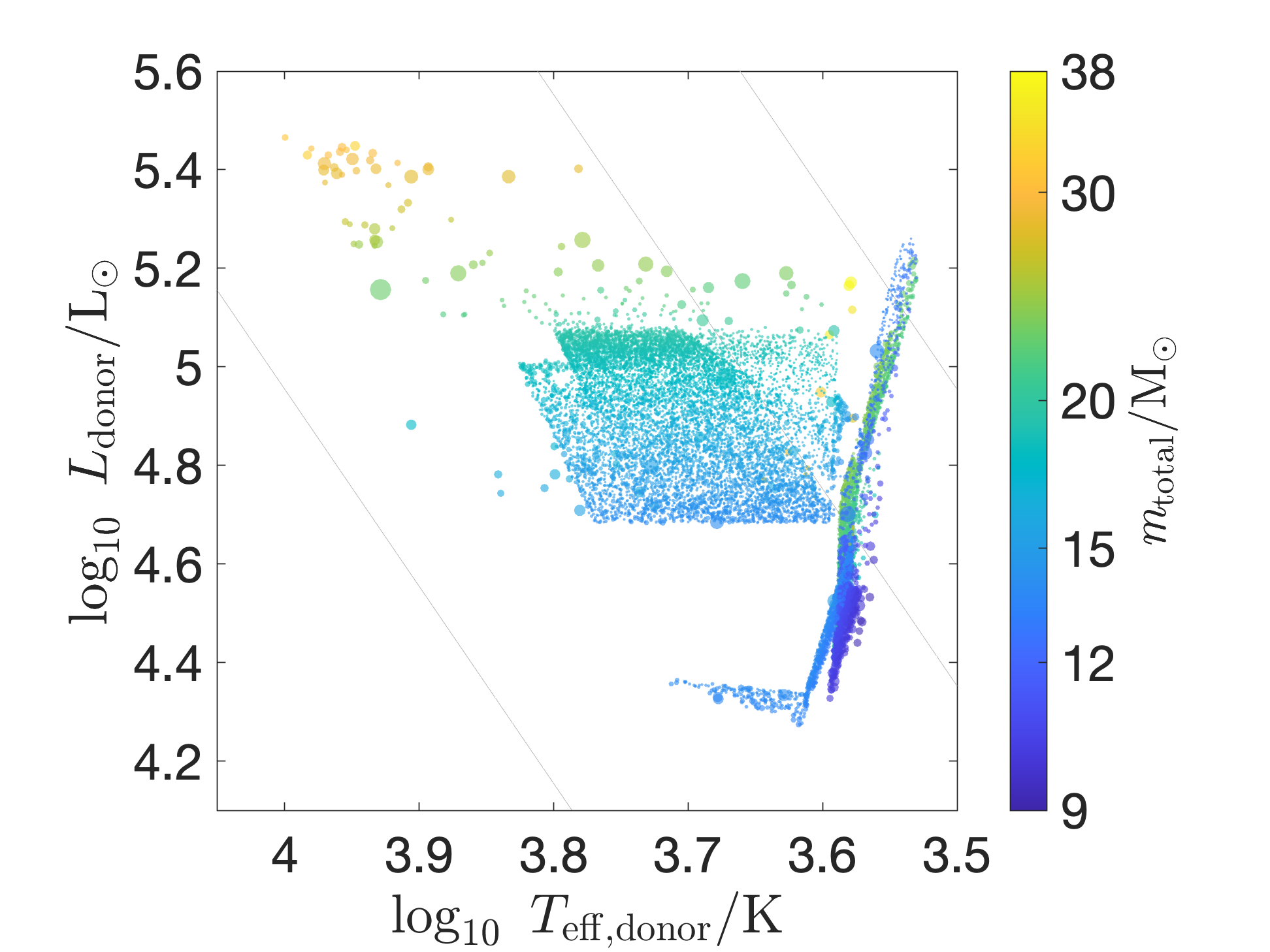}
\includegraphics[width=0.5\textwidth]{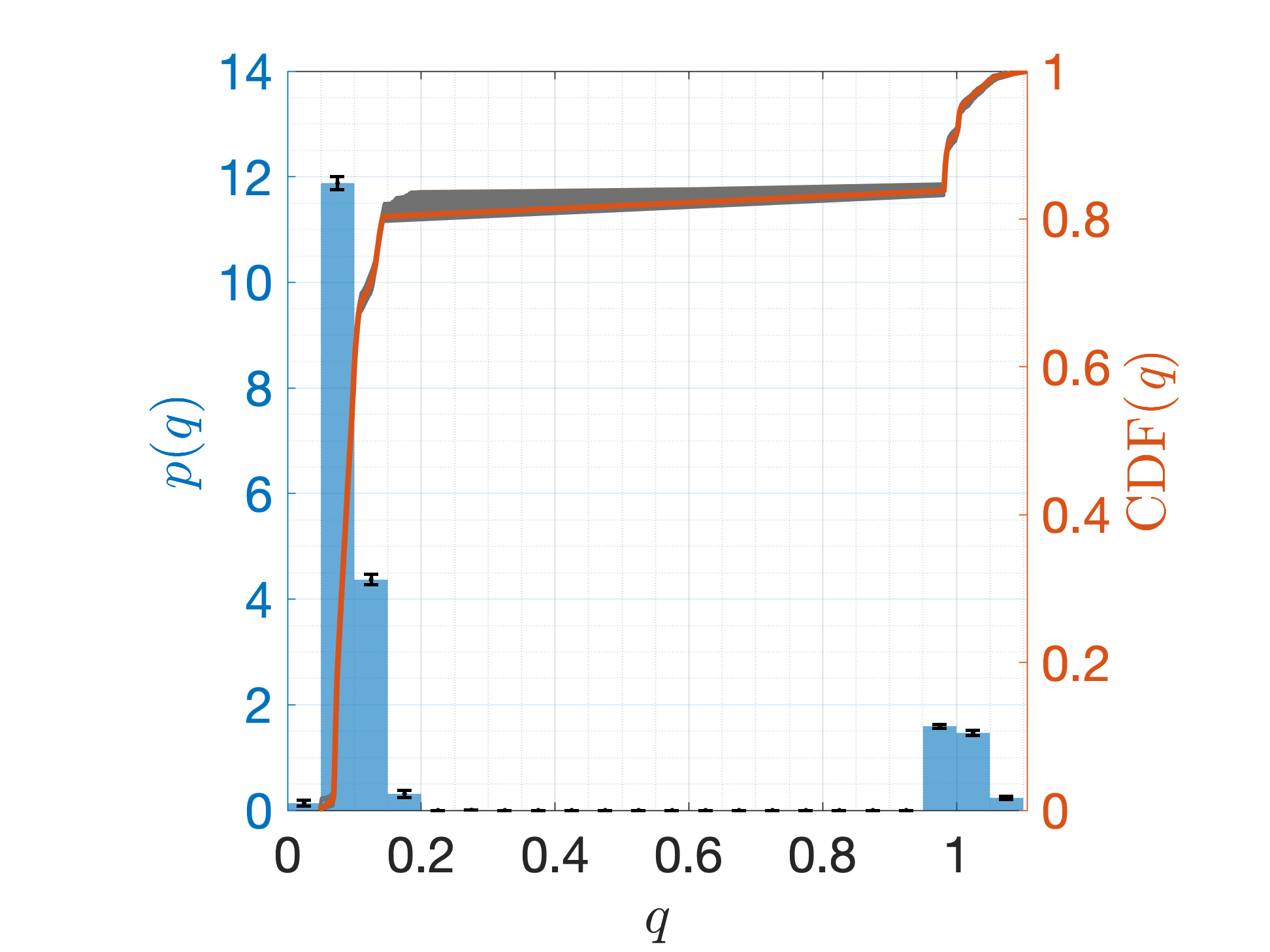}
\includegraphics[width=0.5\textwidth]{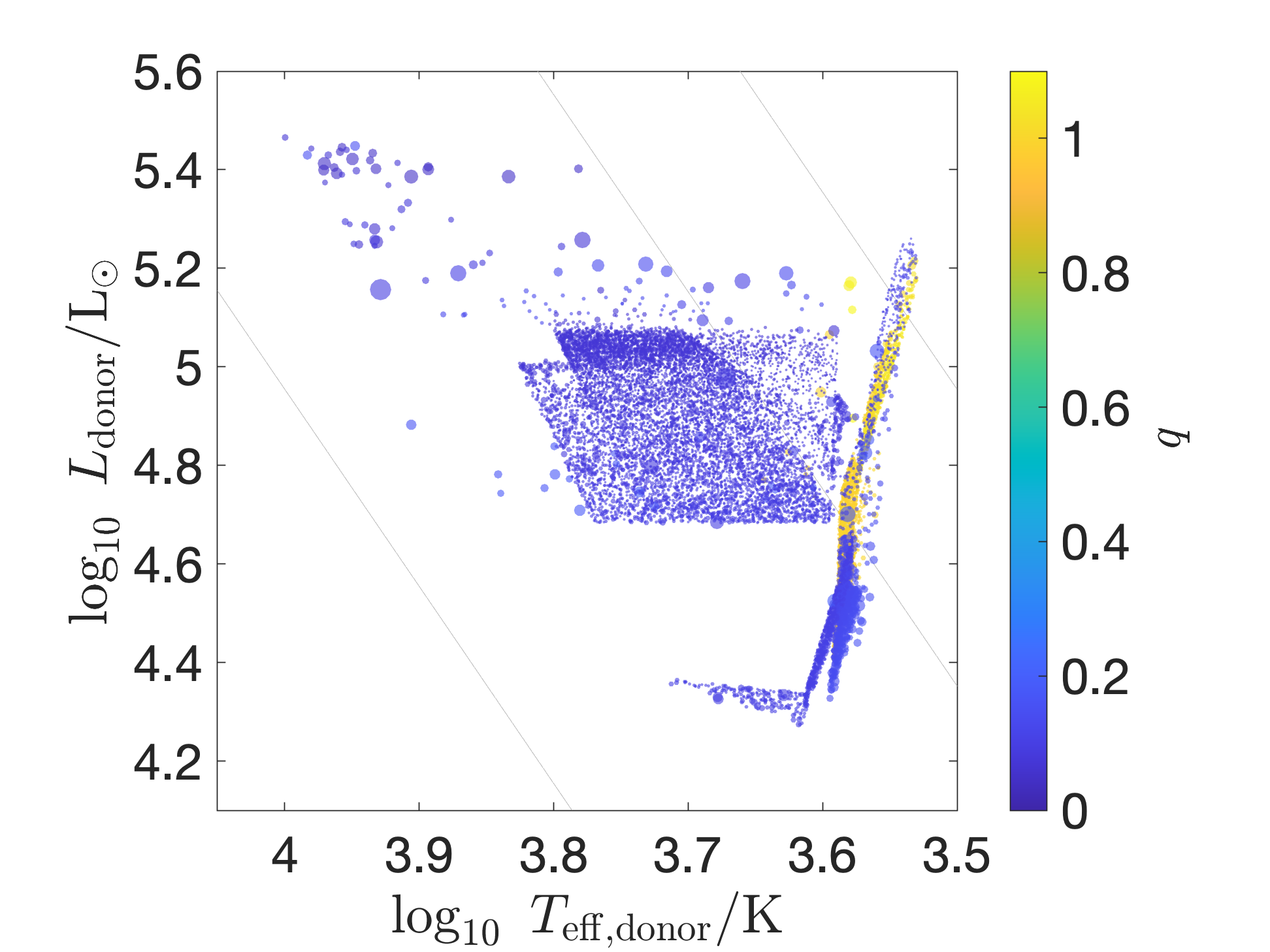}
\caption[Pre-CEE total mass for all DNS-forming systems]{
Pre-CEE mass of all DNS-forming systems. The binary properties presented are total mass (top) and mass ratio (bottom).
For more details, see Section \ref{subsec:propertiesBinary}.
See the caption of Figure \ref{fig:bigHR} for further explanations.
}
\label{fig:massProperties}
\end{figure*}

\subsection{Properties of the Binary}\label{subsec:propertiesBinary}
We also report the properties of each binary by colour coding the property of interest in the \ac{HR} diagram. We report the eccentricity, semi-major axis, total mass ($m_{\rm{total}} = m_{\rm{comp}}+m_{\rm{donor}}$), mass ratio ($q = m_{\rm{comp}}/m_{\rm{donor}}$) and the ratio of the circularisation timescale ($\tau_{\rm{circ}}$) to the radial expansion timescale ($\tau_{\rm{radial}}$), as presented in Table \ref{tab:properties}. All quantities are reported at the onset of the \ac{RLOF} unless stated otherwise.

The eccentricity, semi-major axis and masses of the system determine the orbital energy and angular momentum of a binary (in the two point-mass approximation). The eccentricity and semi-major axis distributions shown in Figure \ref{fig:orbitalProperties} do not account for tidal circularisation.  The eccentricities span the entire allowed parameter range $0 \le e < 1$. The eccentricity distribution has a sharp feature around $e\approx 0$.  Systems with $e\approx 0$ are typically those from Channel II, where the double-core \ac{CEE} happens as the first mass-transfer interaction, without any preceding supernova to make the binary eccentric given our assumption of initially circular binaries (further discussion of this choice is in Section \ref{subsubsection:eccDist}). Meanwhile, the most eccentric binaries have the smallest periapses and interact the earliest during the evolution of the donor, explaining the trend of greater eccentricities for smaller donor sizes in Figure \ref{fig:orbitalProperties}.

The semi-major axis distribution, shown in Figure \ref{fig:orbitalProperties}, has limits of $[a_{\rm{min}},a_{\rm{max}}]=[\lowLimSemiMajorAxis, \uppLimSemiMajorAxis]\ \rm{R_{\odot}}$. The very few extremely wide systems correspond to very eccentric binaries, almost unbound during the supernova explosion (e.g. $e \approx 0.9999$ for the widest binary). While those limits are broad, the limits in periastron are $[a_{\rm{p,min}},a_{\rm{p,max}}] \approx [\lowLimPeriastron,\uppLimPeriastron]\ \rm{R_{\odot}}$. (Very rarely, even smaller periapses are possible when fortuitous supernova kicks send the newly formed \ac{NS} plunging into the envelope of an evolved companion on a very eccentric orbit; however it is not clear whether such events lead to a \ac{CEE} or to a more exotic outcome, such as the formation of a \citealt{ThorneZytkow1977} object). The total mass distribution, shown in Figure \ref{fig:massProperties}, has limits of $[m_{\rm{total,min}},m_{\rm{total,max}}]=[\lowLimTotalMass, \uppLimTotalMass]\ \rm{M_{\odot}}$.

We compute the mass ratio at the onset of the \ac{RLOF} leading to the \ac{CEE}. The mass ratio, shown in Figure \ref{fig:massProperties}, has limits of $[q_{\rm{min}},q_{\rm{max}}]=[\lowLimMassRatio, \uppLimMassRatio]$. The broad distribution in fact consists of two distinct peaks, one close to $q=0$ and the other close to $q=1$, with a large gap between $0.18 \leq q \leq 0.98$ (see Figure \ref{fig:massProperties}).
The extreme mass ratio systems correspond to \acp{CEE} from Channel I, where the companion is a \ac{NS}. The $q \approx 1$ systems correspond to \acp{CEE} from Channel II, where there is a double-core \ac{CEE} with a non-compact companion star. 
The systems with $q > 1$ are double-core \ac{CE} systems with $q_{\rm{ZAMS}} \approx 1$ which, at high metallicity, may reverse their mass ratio via mass loss through winds before the primary star expands and undergoes \ac{RLOF}.

\subsection{Tidal Timescales in Pre-Common-Envelope Systems}
\label{subsec:TidalTimescales}
Given the uncertainties in the treatment of tides, and our interest in comparing the impact of different tidal prescriptions as discussed below (see Section \ref{subsec:tidalUncertainty}), we do not include tidal synchronisation or circularisation in binary evolution modelling for this study. Instead, we consider whether tides would be able to efficiently circularise the binary before the onset of \ac{RLOF} leading to a \ac{CEE}. As discussed in Section \ref{subsec:radialExpansion}, we use the ratio of the circularisation timescale to the radial expansion timescale as a proxy for the efficiency of tidal circularisation of an expanding star about to come into contact with its companion. If $\tau_{\rm{circ}}/\tau_{\rm{radial,donor}} > 1$, we label the binary as still eccentric at \ac{RLOF}.  Given that the circularisation timescale is longer than the synchronisation timescale (see Section \ref{subsec:tides}), we focus on the former and assume that if the binary is able to circularise, it will already be synchronous.  

\begin{figure*}
\includegraphics[width=0.5\textwidth]{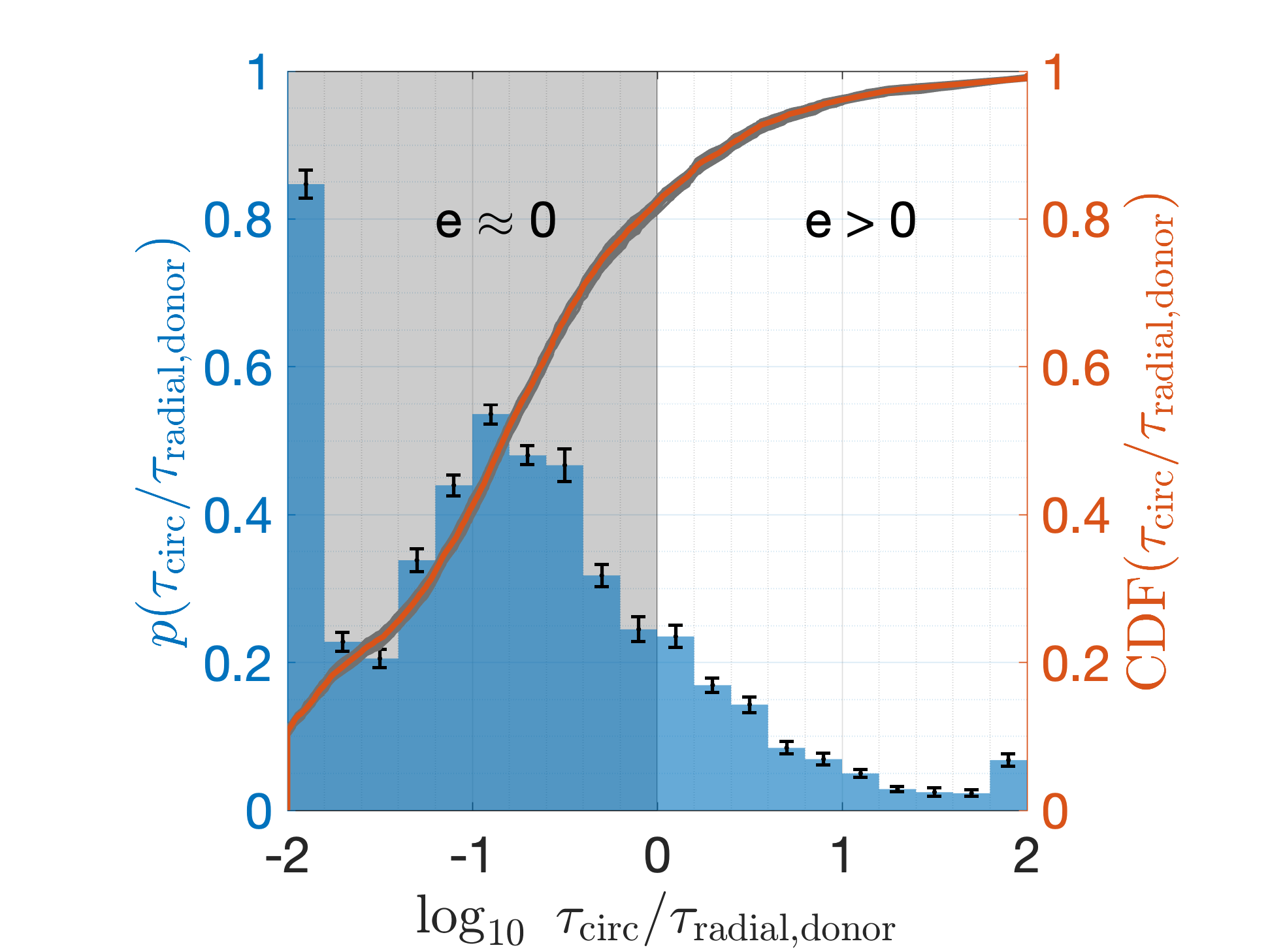}
\includegraphics[width=0.5\textwidth]{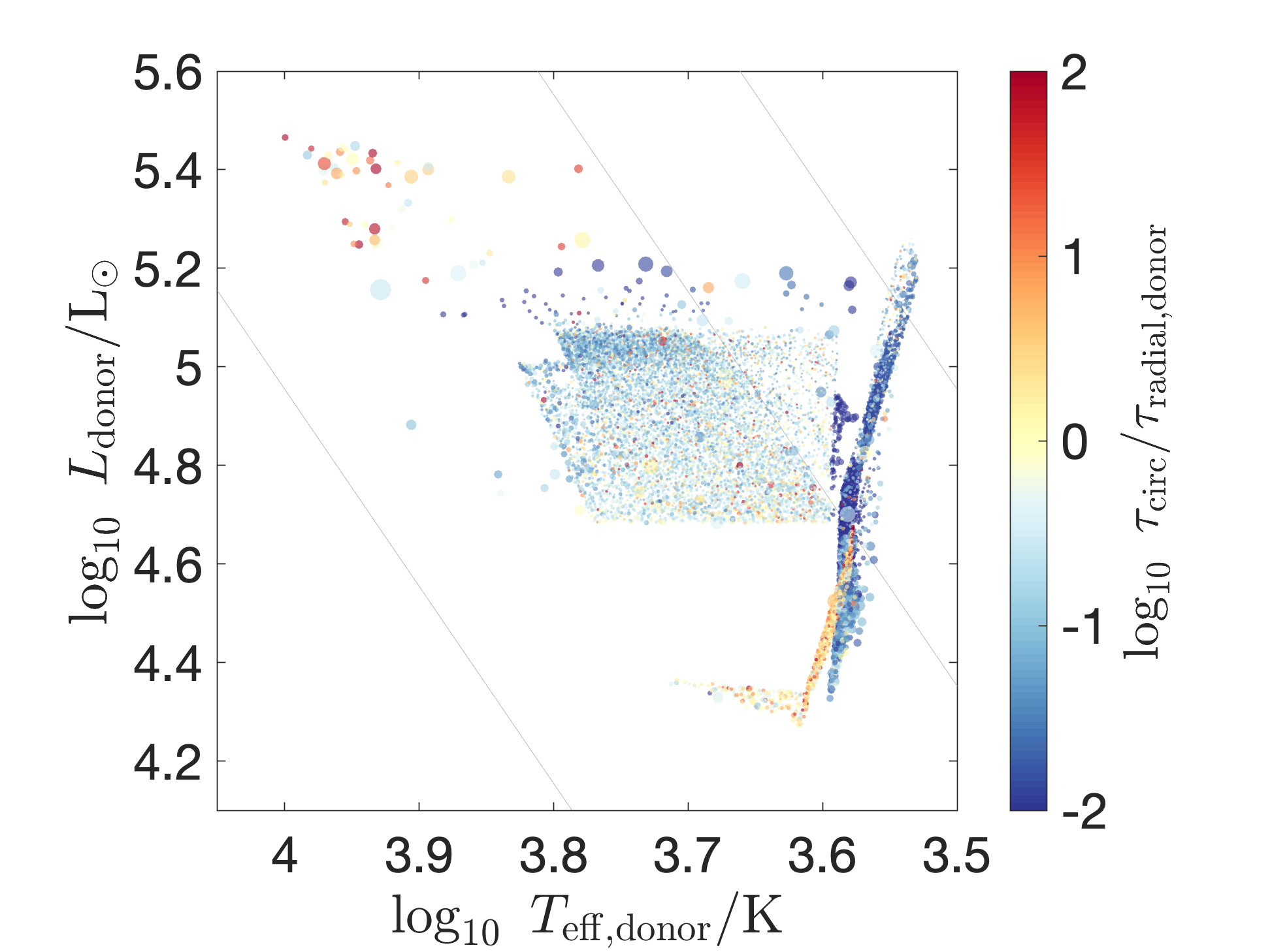}
\caption{
Ratio of tidal circularisation timescale to the star's radial expansion timescale for all DNS-forming systems. We present the default scenario where all evolved stars, including \ac{HG} and \ac{CHeB} stars, are assumed to have formed a fully convective envelope. If $\log_{10}(\tau_{\rm{circ}}/\tau_{\rm{radial}})\le 0$, we assume that binaries circularise before the onset of the \ac{CEE}. Binaries indicated with blue (red) dots are predicted to have circular (eccentric) orbits.
We cap $\ -2 \le \log_{10}(\tau_{\rm{circ}}/\tau_{\rm{radial}}) \le 2$ to improve the plot appearance. The grey shaded region in the histogram highlights the systems which circularise by the onset of \ac{RLOF}. 
For more details, see Section \ref{subsec:TidalTimescales}.
See the caption of Figure \ref{fig:bigHR} for further explanations. 
}
\label{fig:convectiveTidalTimescales}
\end{figure*}

Figure \ref{fig:convectiveTidalTimescales} shows the ratio $\tau_{\rm{circ}}/\tau_{\rm{radial,donor}}$ under our default assumption in which both \ac{HG} and \ac{CHeB} stars have fully convective envelopes for the purpose of tidal circularisation calculations and experience the equilibrium tide. This assumption results in \howManyCircConv\% of the systems being circular at the onset of \ac{RLOF}.

The analysis of circularisation timescales is mostly relevant for systems formed through Channel I (see Section \ref{subsec:tidalUncertainty}). There are two reasons for this. The first one is that they are expected to acquire a non-zero eccentricity after the first supernova. The second one is that they are more likely to have a radiative or only partially convective envelope, making circularisation less efficient. On the other hand, systems formed through Channel II have a fully-convective envelope, which allows for efficient tidal circularisation and synchronisation.  We apply the low-eccentricity approximation described in Section \ref{subsubssec:Hut} to computing the tidal timescales of these systems, even though they are circular by construction before the first \ac{SN} (we discuss this assumption in Section \ref{subsubsection:eccDist}).

\section{Discussion}
\label{sec:discussion}
In this section we discuss the properties of \acp{CEE} experienced by isolated stellar binaries evolving into \acp{DNS} and present some of the caveats in our COMPAS rapid population synthesis models.

\subsection{Common-Envelope Episode Sub-populations in Evolving Double Neutron Stars} 
\label{subsec:subpopulations}
\subsubsection{Formation channels}
There are two main formation channels leading to \ac{DNS} formation. Channel I involves high mass ratio single-core \ac{CEE} between
a \ac{NS} primary and a post-\ac{MS} secondary. Channel I  has been studied thoroughly in the literature, e.g. \cite{bhattacharya1991formation,tauris2006formation,tauris2017formation} and references therein.
Channel II involves a double-core common envelope between two post-\ac{MS} stars.  A similar channel has been proposed by \cite{brown1995doubleCore} and \cite{dewi2006double}, among others. 
Channel II requires similar masses at \ac{ZAMS} driven by the need of similar evolutionary timescales so that both stars are post-\ac{MS} giants at the time of their first interaction. For low (high) mass stars, the difference in \ac{ZAMS} mass can be up to 3 (7)\%, in agreement with \cite{dewi2006double}. Our Channel II has an additional case BB mass transfer episode from a helium-shell-burning primary onto a helium-\ac{MS} secondary.

\subsubsection{Sub-populations and tidal circularisation}
\begin{figure*}
\includegraphics[width=0.5\textwidth]{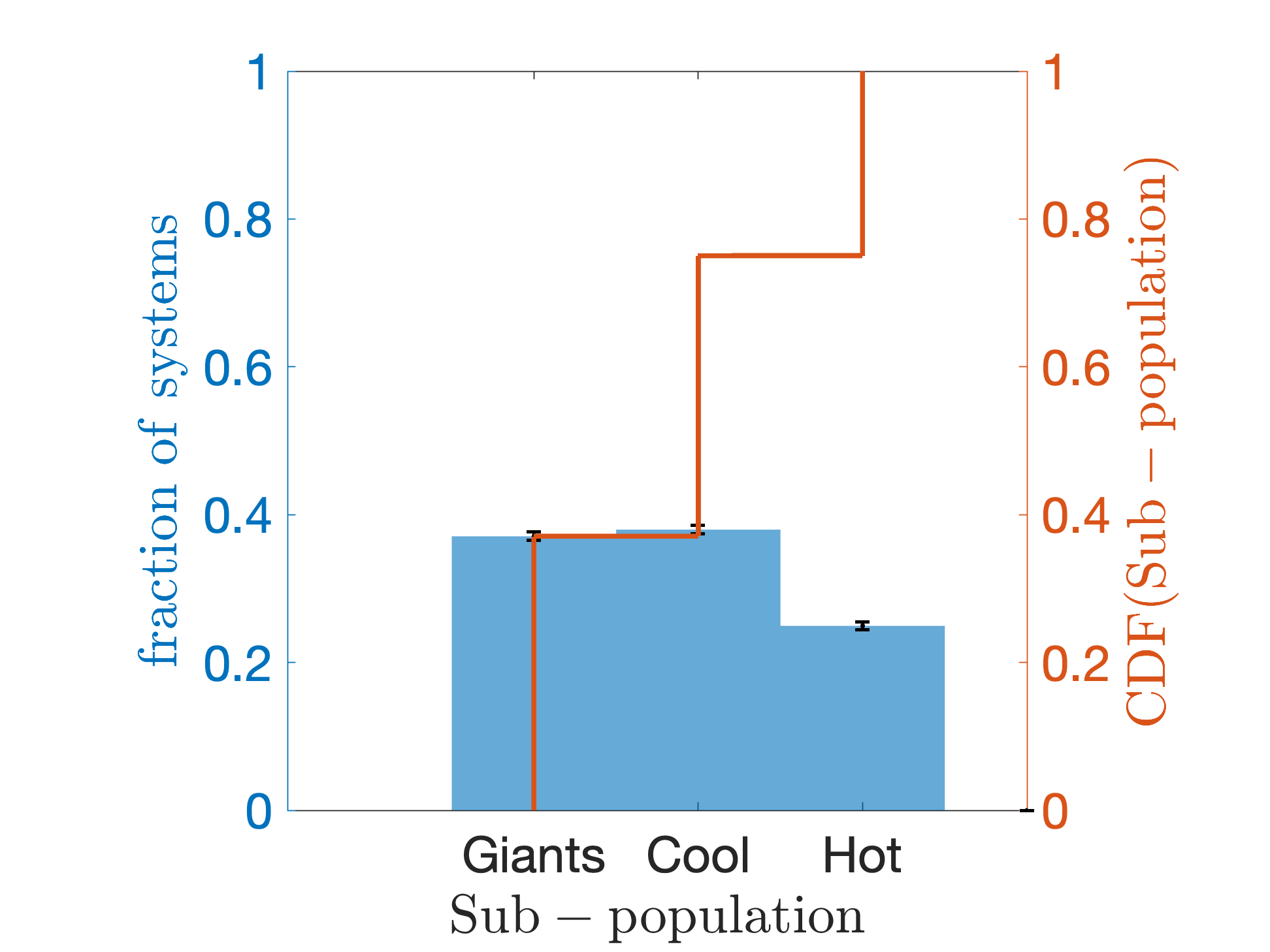}
\includegraphics[width=0.5\textwidth]{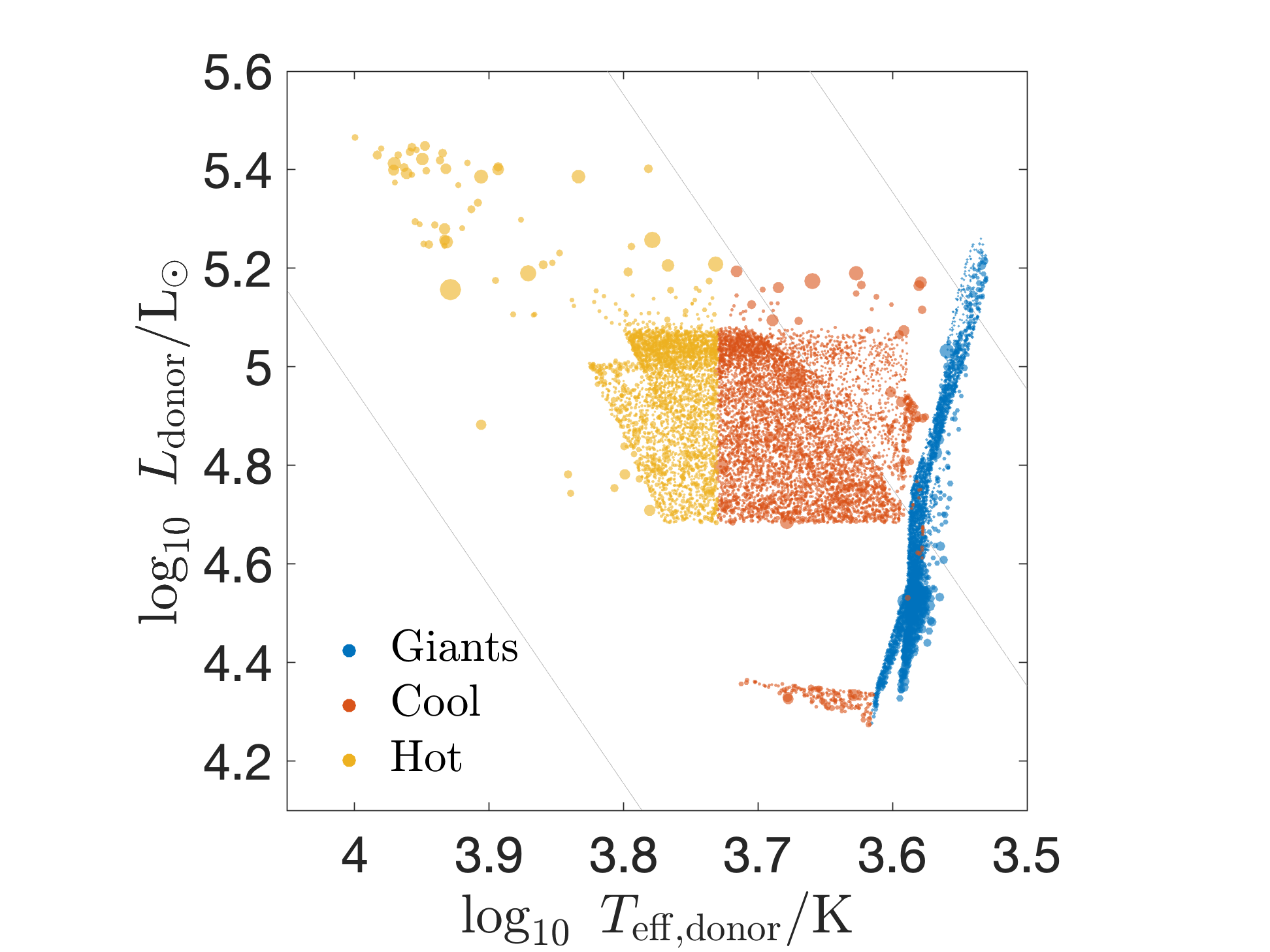}
\caption{
\ac{DNS}-forming binaries clustered by the donor type at the onset of the \ac{CEE}.
Sub-populations: (a) giant donors with fully-convective envelopes in blue, (b) \ac{HG} or \ac{CHeB} donors with partially-convective envelopes in red, and (c) \ac{HG} or \ac{CHeB} donors which have not yet formed a deep convective envelope in yellow.
For more details, see Section \ref{subsec:propertiesDonor}.
See the caption of Figure \ref{fig:bigHR} for further explanations.
}
\label{fig:threePopulations}
\end{figure*}

We separate \ac{CEE} donors into three distinct sub-populations depending on their evolutionary phase at the onset of \ac{RLOF}: \textit{giants}, \textit{cool} and \textit{hot} (see Table \ref{tab:subpopulation} and Figure \ref{fig:threePopulations}).   

The first one, giants, correspond to giant donors with fully-convective envelopes. The other two sub-populations correspond to \ac{HG} or \ac{CHeB} donors, most of them evolving via the single-core Channel I. We distinguish between cool donors with a partially convective envelope and hot donors with a radiative envelope. We follow \citet{belczynski2008compact} in using the temperature $\log_{10}\ (T_{\rm{eff,donor}}/\rm{K}) = 3.73$ as the boundary between the cool and hot sub-populations.

The presence and depth of a convective envelope impacts the response of the star to mass loss and, hence, the dynamical stability of mass transfer.  In particular, hot donors lacking a deep convective envelope may be stable to mass transfer and avoid a \ac{CEE}.  At the same time, some of the less evolved hot donors may not survive a \ac{CEE} even if they do experience dynamical instability (pessimistic variation). 

\cite{Klencki2020bindingEnergy} use detailed stellar evolution models to argue that only red supergiant donors with deep convective envelopes are able to engage in and survive a \ac{CEE}. For their assumptions, this would reduce the estimated rate of \ac{DNS} formation. However, and similar to this study, they focus on \ac{RLOF} structures and not the structures at the moment of the instability.

\begin{figure}
\includegraphics[trim={1.cm 0.0cm 1.cm 0.0cm},clip,width=0.48\textwidth]{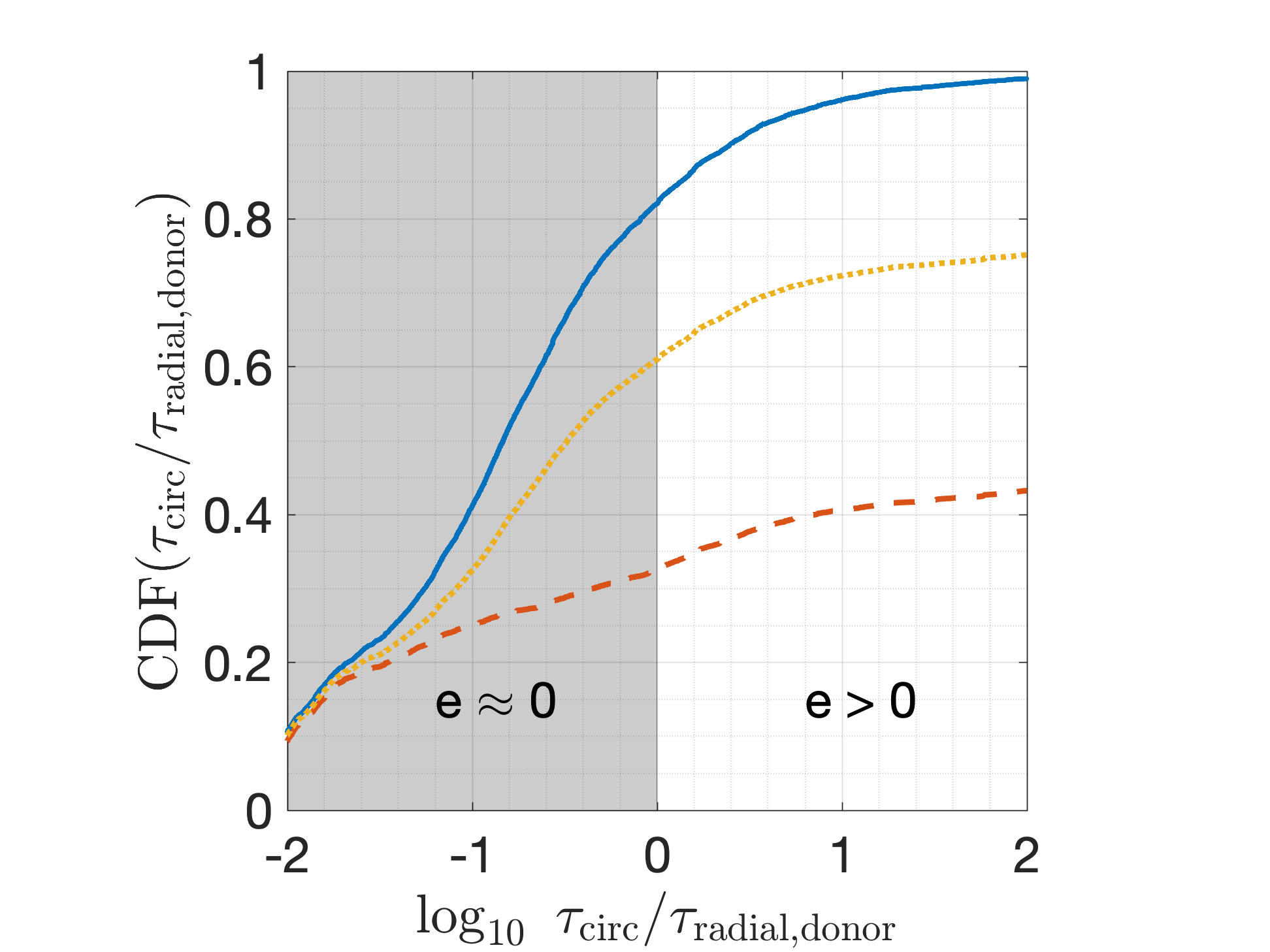}
\caption{
\ac{CDF} of the ratio of the circularisation timescale to the donor radial expansion timescale computed at \ac{RLOF} onset leading to \ac{CEE} for all \ac{DNS}-forming systems. Here we present three scenarios. The solid blue line is our default assumption: all donors have a deep convective envelope (same as in left panel of Figure \ref{fig:convectiveTidalTimescales}). The red dashed line follows \cite{hurley2002evolution} with the assumption that \ac{CHeB} tidal evolution is dominated by the dynamical tide,  i.e. that \ac{CHeB} stars have a radiative envelope. The yellow dotted line follows \cite{belczynski2008compact} in assuming that stars with $\log\ T_{\rm{eff}}\le 3.73\ \rm{K}$ have a fully convective envelope, for both \ac{HG} and \ac{CHeB} donors; and a fully radiative envelope otherwise, as in Figure \ref{fig:threePopulations}. 
For more details, see Section \ref{subsec:TidalTimescales}. 
}
\label{fig:comparisonCircularisationTimescales}
\end{figure}

Here, we focus only on the impact of the assumed structure of the donor on the efficiency of tidal circularisation and do not account for possible consequences for mass transfer stability.  We compare three alternative models in Figure \ref{fig:comparisonCircularisationTimescales}.
 
Our default tidal circularisation model assumes that all evolved donors, including both \ac{HG} and \ac{CHeB} stars,  have fully convective envelopes, and therefore experience efficient equilibrium tides. 
Our default assumption estimates that $\howManyEccConv$\% of systems will be eccentric at the onset of the \ac{RLOF} leading to the \ac{CEE}. 
This is the lowest fraction of eccentric systems among all variations because tides are particularly efficient for stars with convective envelopes.

In reality, \ac{CHeB} stars are expected to begin the \ac{CHeB} phase with a radiative envelope and develop a deep convective envelope by the end of it. The single stellar fits from \citet{hurley2000comprehensive} do not contain explicit information about the moment when this transition occurs. \citet{hurley2002evolution} assume that all \ac{CHeB} stars have a radiative envelope and that the dynamical tide is dominant in their tidal evolution. Adopting this assumption leads to $\howManyEccRad$\% of binaries remaining eccentric at the onset of the \ac{RLOF} leading to the \ac{CEE}.

Alternatively, \citet{belczynski2008compact} assume that hot stars with $\log_{10}\ (T_{\rm{eff}}/\rm{K}) > 3.73$ have a radiative envelope, while cool stars with $\log_{10}\ (T_{\rm{eff}}/\rm{K}) \le 3.73$ have a convective envelope. Adopting this assumption leads to $\howManyEccBelc$\% of binaries remaining eccentric at the onset of the \ac{RLOF} leading to the \ac{CEE}.

According to our estimates, a significant fraction of systems will be eccentric at \ac{RLOF}. These estimates were made within the framework of the fitting formulae for single stellar evolution from \cite{hurley2000comprehensive}. More detailed fitting formulae, which include the evolutionary stage of stars as well as the mass and radial coordinates of their convective envelopes, would allow for a self-consistent determination of whether a star has a radiative, a partially convective or a fully convective envelope for both dynamical stability and tidal circularisation calculations. 


\begin{table*}
\caption{Distinct \ac{DNS} sub-populations as described in Section \ref{subsec:subpopulations} and presented in Figure \ref{fig:threePopulations}.}
\centering
\begin{tabular}{@{}ccccccc@{}}
\hline\hline
Sub-population & Threshold & Dominant Channel & Donor & Envelope & Colour & Fraction \\
\hline%
Giants & - & II (double core) & \ac{GB}, \ac{EAGB} & fully convective & blue &  \fractionGiants \\
Cool & $\log_{10} (T_{\rm{eff}}/\rm{K}) < 3.73$ & I (single core) & \ac{HG}, \ac{CHeB} & partially convective & orange & \fractionCool\\
Hot & $\log_{10} (T_{\rm{eff}}/\rm{K}) \ge 3.73$ & I (single core) & \ac{HG}, \ac{CHeB} & radiative/convective & yellow & \fractionHot \\
\hline\hline
\end{tabular}
\label{tab:subpopulation}
\end{table*}

\begin{figure}
\includegraphics[trim={1.cm 0.0cm 1.cm 0.0cm},clip,width=0.48\textwidth]{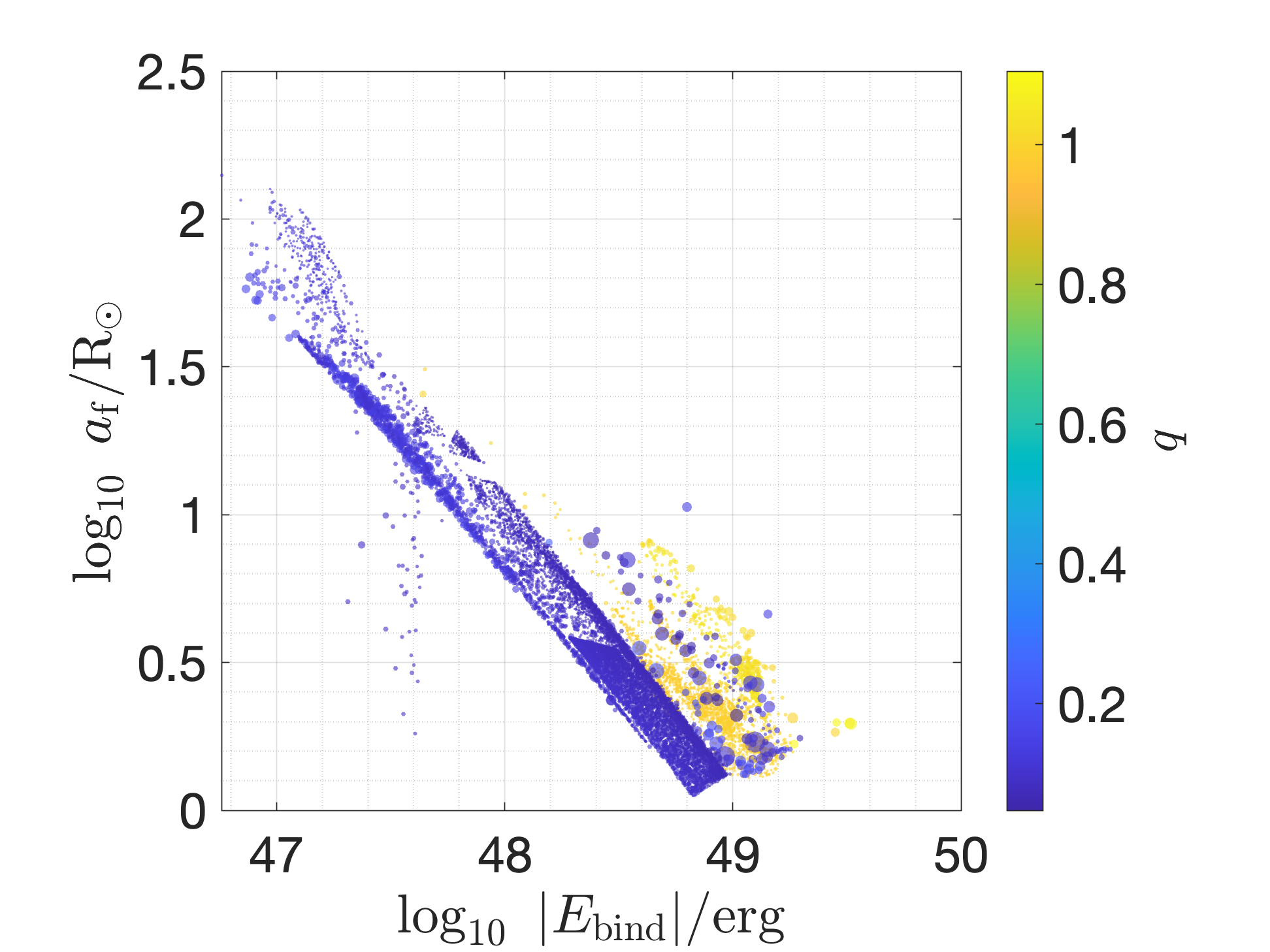}
\caption{
All \ac{DNS}-forming binaries from our Fiducial model are shown here. 
We present the post-\ac{CEE} separation $a_{\rm{f}}$ as a function of the absolute value of the envelope binding energy $|E_{\rm{bind}}|$.
For the double-core scenario, the binding energy is $E_{\rm{bind}}=E_{\rm{bind,donor}}+E_{\rm{bind,comp}}$. 
The size of the marker indicates the sampling weight and its colour shows the mass ratio $q$.
This Figure can be compared to Figures 1 and 2 from \cite{IaconiDeMarco2019}. 
That study presents simulations of \ac{CE} binaries and observations of post-\ac{CE} binaries.
Most systems presented here do not feature in \cite{IaconiDeMarco2019}.
}
\label{fig:Orsola}
\end{figure}

\begin{figure*}
\includegraphics[width=0.5\textwidth]{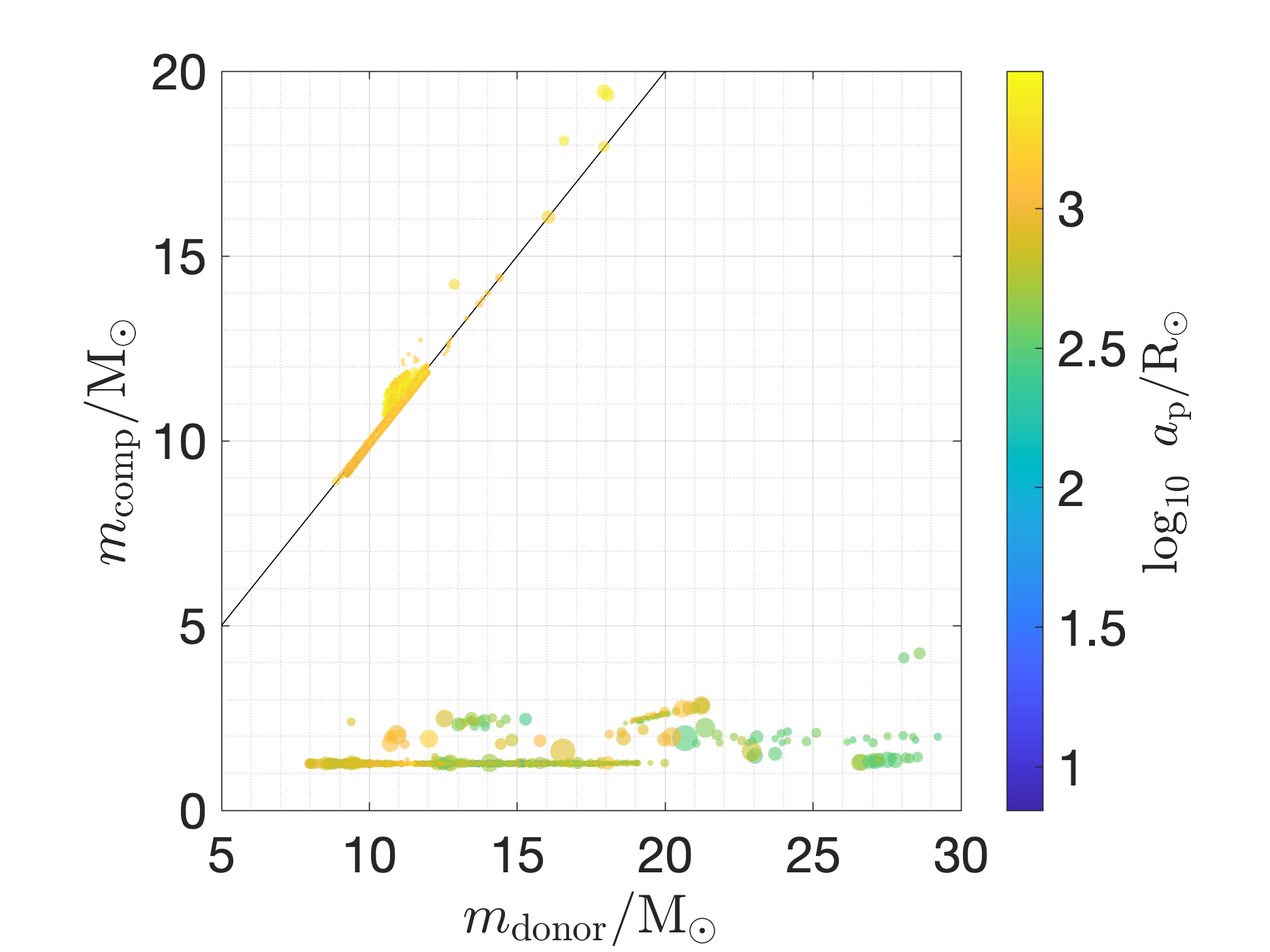}
\includegraphics[width=0.5\textwidth]{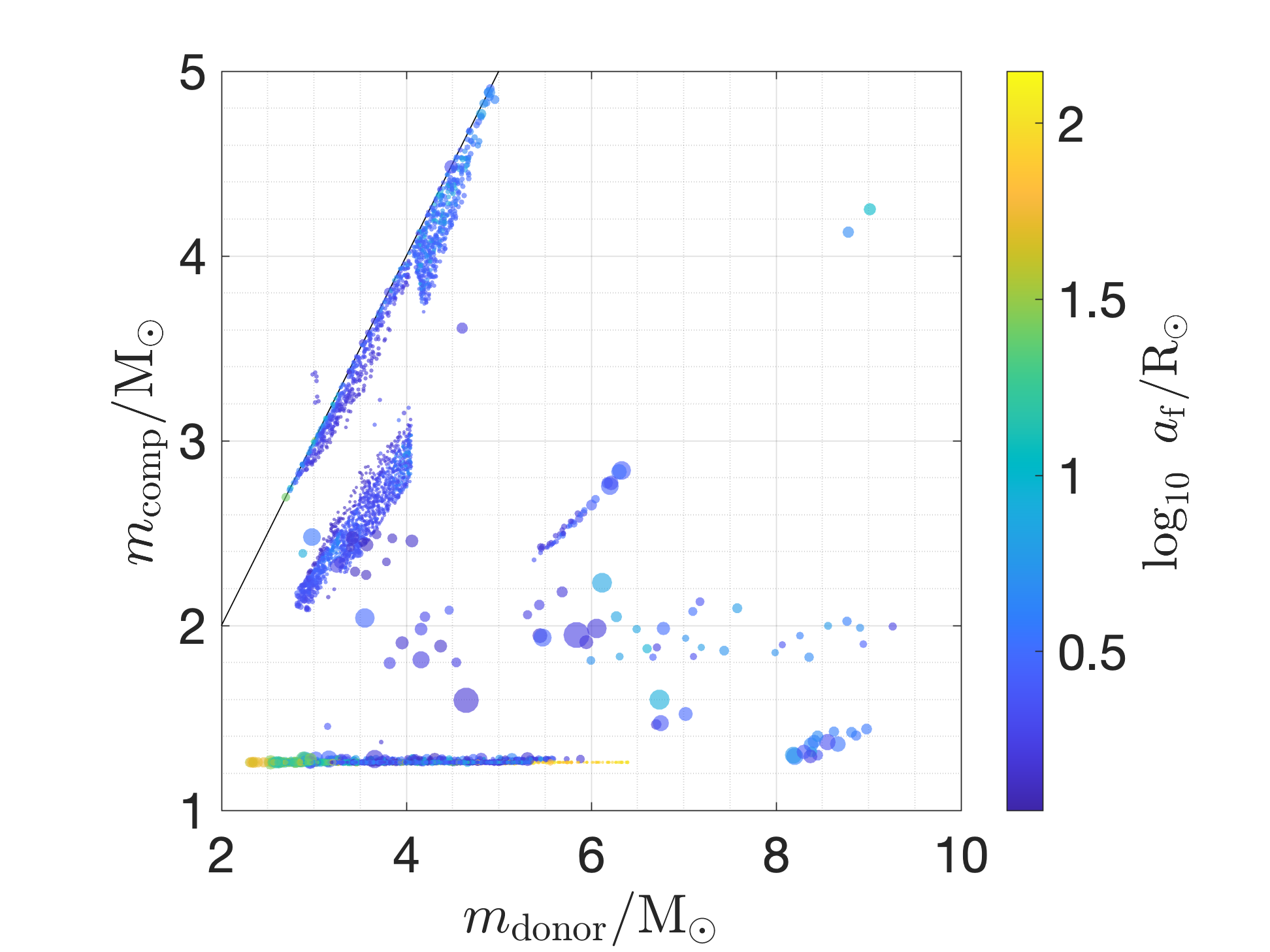}
\caption{
Binary separations at \acp{CEE} leading to \acp{DNS} at the onset of \ac{RLOF} (left) and after the \ac{CEE} (right). We show the donor ($m_{\rm{donor}}$) and companion ($m_{\rm{comp}}$) mass in both plots, with a solid grey line indicating $m_{\rm{donor}}=m_{\rm{comp}}$. The colour bars, with different scales, show the pre-\ac{CEE} periastron (left) and final semi-major axis (right).
}
\label{fig:massesAndSeparations}
\end{figure*}

\subsection{Common-Envelope Episodes as candidates for luminous red novae transients} 
\label{subsec:LRNe}
Recently, the luminous red nova transient \LRNevent\ was reported by \citet{Blagorodnova2017LRN}. This event is similar to other luminous red novae associated with \acp{CEE} \citep{Ivanova2013}. Following the discovery of \LRNevent, archival photometric data from earlier epochs were found. \citet{Blagorodnova2017LRN} used these to derive the characteristics of the progenitor. The inferred properties of the progenitor of \LRNevent\ are a luminosity of $L_{\rm{donor}}\approx 87,000\ \rm{L_{\odot}}$, an effective temperature of $T_{\rm{eff,donor}}\approx 7,000\ \rm{K}$ and a mass of $m_{\rm{donor}}=18 \pm 1\ \rm{M_{\odot}}$ (see Figure \ref{fig:bigHR} for location in the \ac{HR} diagram). 

\cite{Blagorodnova2017LRN} found that the immediate pre-outburst progenitor of \LRNevent\ was consistent with an F-type yellow supergiant crossing the HG. If we take the inferred values for this star as the values at the onset of \ac{RLOF}, then this star is consistent with pre-\ac{CEE} stars in our predicted distribution of \ac{DNS}-forming systems. However, we emphasise that the appearance of the donor star can change significantly between the onset of \ac{RLOF}, i.e., the point at which the models shown in Figure \ref{fig:bigHR} are plotted, and dynamical instability.

\cite{Howitt2019LRNe} explored population synthesis models of luminous red novae.  
Here we use the same pipeline adopted for that study to explore the connection to \ac{DNS} populations.
Doing so, fewer than 0.02\% of all luminous red novae lead to \acp{DNS}. These are amongst the most energetic luminous red novae and would be over-represented in the magnitude-limited observable population. 
Future \acp{DNS} constitute nearly 10\% of the subpopulation of luminous red novae with predicted plateau luminosities greater than $10^7\ \rm{L_{\odot}}$.  

\subsection{Eccentric Roche-lobe Overflow leading to a Common-Envelope Episode}
We predict that the sub-population of giant donors with fully-convective envelopes and cool donors with partially-convective envelopes are likely to be circular at the onset of the \ac{CEE} (see Figures \ref{fig:convectiveTidalTimescales} and \ref{fig:comparisonCircularisationTimescales}).  On the other hand, we find that the sub-population of hot donors often does not circularise by the onset of the \ac{CEE}. This sub-population with hot donors are binaries with high eccentricities at the onset of the \ac{RLOF} (see Figure \ref{fig:orbitalProperties}).

This result raises questions about the initial conditions of a \ac{CEE}, which is often assumed to begin in a circular orbit, both in population synthesis studies and in detailed simulations.  Population synthesis codes such as SEBA \citep{portegies1996SEBA,portegies1998formation,toonen2012SEBA}, \texttt{STARTRACK} \citep{belczynski2002comprehensive,belczynski2008compact}, BSE \citep{hurley2002evolution}, the Brussels code \citep{de2004influence}, COMPAS \citep{stevenson2017formation,VignaGomez2018DNSs}, ComBinE \citep{Kruckow2018ComBinE} and customised software based on them all assume that \ac{RLOF} commences in circular binaries. Detailed simulations, such as those of \citet{Passy2012CEE}, \cite{MacLeod2018CEEs} and others, often make the assumption of an initially circular orbit (but see \citealt{Staff2016}, discussed in more detail in Section \ref{sec:eccMT}). 

\subsubsection{Theory of mass transfer in eccentric binaries}
Mass transfer in eccentric binaries has been explored with both semi-analytical and analytical methods \citep{MateseWhitmire1983a,MateseWhitmire1983b,Sepinsky2007,Sepinsky2009,Sepinsky2010,Dosopoulou2016I,Dosopoulou2016II}. 

The analysis of \cite{Sepinsky2007} et al. assumes fully conservative mass transfer.  While they consider mass transfer from a stellar donor onto a neutron star, this assumption and the typical $10^{-9}\ \rm{M_{\odot}\ yr^{-1}}$ mass transfer rate they consider is relevant for low-mass X-ray binaries, not \ac{DNS} progenitors as discussed here.

\cite{Dosopoulou2016II} study orbital evolution considering both conservative and non-conservative mass transfer.  The latter scenario is particularly relevant for \ac{DNS} formation.  We assume a mass transfer rate of $10^{-5}\ \rm M_{\odot}\ yr^{-1}$, a $1.44\ \rm M_{\odot}$ \ac{NS} and mass loss from the vicinity of the \ac{NS} (isotropic re-emission) as parameters in their Equation (44).  Highly eccentric systems ($e>0.9$) have circularisation timescales of more than 1 Myr due to mass transfer, under their assumption that all mass transfer happens at periapsis.  This timescale is reduced to around a thousand years for mass transfer in $e<0.1$ binaries, although this can be very sensitive to assumptions about the specific angular momentum lost at the level of an order of magnitude. The assumption of instantaneous mass transfer at periapsis is questionable for low-eccentricity binaries, precisely those which may efficiently circularise through mass transfer.

\cite{HamersDosopoulou2019ApJ} noted that evolution towards circularisation from \cite{Dosopoulou2016II} could lead to (nonphysical) negative eccentricity solutions. 
They proposed a revised analytic model for mass transfer in eccentric binaries. This study takes into account the separation and eccentricity evolution of an initially eccentric system at \ac{RLOF}. However, their model is only valid in the regime of fully-conservative mass transfer, and is therefore more restricted than the general formalism \cite{Dosopoulou2016II}. Mass transfer episodes in binaries which will become \acp{DNS} are typically non-conservative. Mass transfer from a post-\ac{MS} donor onto a \ac{MS} companion, such as the first mass transfer episode from Channel I, is generally only partly conservative \citep{schneider2015evolution}. Mass transfer onto a \ac{NS} companion is highly non-conservative, almost in the fully non-conservative limit \citep{tauris2015ultra}.

A full understanding of the evolution of eccentric systems in \ac{RLOF} is yet to be achieved.
A detailed treatment of non-conservative mass transfer in an eccentric binary could yield different criteria for dynamical stability and, ultimately, for determining if a system engages in a \ac{CEE}.

\subsubsection{Modelling of mass transfer in eccentric binaries}\label{sec:eccMT}
Numerical methods and simulations have also been used to study mass transfer in eccentric binaries \citep{Regos2005,Church2009,LajoieSills2011,vanDerHelms2016, Staff2016,Bobrick2017}.

\cite{Staff2016} carried out hydrodynamic simulations of a $\approx 3\ \rm{M_{\odot}}$ giant star with a less massive \ac{MS} companion in an eccentric orbit. They conclude that eccentric systems transfer mass only during the periastron passage, which delays the onset of the \ac{CEE}. Each periastron passage also makes the binary less eccentric.

\cite{Gilkis2019} discuss the passage of a \ac{NS} through the envelope of a giant star, likely on an eccentric orbit, and conclude that the system might be able to eject the envelope or lead to a merger between the neutron star and the core, which they call a \ac{CE} jets supernova \citep[][see also \citealt{Schroeder:2019}]{SokerGilkis2018}. The former results in a less luminous transient, comparable in energetic to luminous red novae \citep{KashiSoker2016}.
The interaction is driven by jets which might enhance mass loss at periastron passages, keeping the system eccentric \citep{KashiSoker2018}. These jets might, in some cases, prevent \acp{CEE} \citep{ShiberSoker2018}.

\subsubsection{Observations of eccentric mass-transferring binaries}
Eccentric semi-detached and contact binaries, i.e. mass transferring binaries, have been previously observed in low-mass systems \citep{PetrovaOrlov1999}. Eccentric ($e\lesssim 0.2$) \ac{MS}-White Dwarf binaries which are believed to have experienced \ac{RLOF} are not rare \citep{Vos2013,Kawahara2018,Masuda2019}. \cite{Jayasinghe2019} found a more massive B-type Heartbeat star in an eccentric ($e=0.58$) orbit. Heartbeat stars exhibit clear signatures of tidal oscillations at each periastron passage. While there is no evidence for accretion, it is likely that the system reported in \cite{Jayasinghe2019} will engage in \ac{RLOF} at some later point .

Sirius \citep{vandenBos1960Sirius,Gatewood1978Sirius} is a \ac{MS}-White Dwarf binary with $e=0.59$ which, according to canonical binary evolution dynamics, should have circularised when the White Dwarf progenitor became a giant. \cite{BonacicMarinovic2008} propose a model which allows for tides, mass loss and mass transfer in an eccentric orbit, physically motivated by Sirius. Following \cite{BonacicMarinovic2008}, \cite{SaladinoPols2019} carried out hydrodynamic simulations of binary stars with significant wind-driven mass loss and find that this eccentricity-enhancing mechanism is non-negligible.

\subsubsection{$\gamma ^2$ Velorum as an eccentric massive post-mass-transferring binary}
\cite{North2007} reported the orbital solution and fundamental parameter determination of the massive binary $\gamma^2$ Velorum. This binary has a reported period of $78.53\pm0.01$ days and an eccentricity of $0.334\pm0.003$, with an inferred mass of $28.5\pm1.1\ \rm{M_{\odot}}$ for the O-star primary and $9.0\pm0.6\ \rm{M_{\odot}}$ for the Wolf-Rayet secondary. While $\gamma^2$ Velorum did not experience a \ac{CEE}, it could have experienced some mass transfer as an eccentric system. \cite{Eldridge2009} discusses $\gamma^2$ Velorum as post-mass-transfer binary system. In that work, \cite{Eldridge2009} takes into account how the evolutionary stage of the donor during mass transfer determines the efficiency of tidal circularisation. They point out that a less evolved star with a radiative envelope is not likely to circularise during the mass transfer phase. This would lead to a post-mass-transfer eccentric system such as $\gamma^2$ Velorum.


\subsection{Caveats and Limitations}
\subsubsection{\ac{CEE} and delayed dynamical instability} 
\label{subsubsec:delayed}
The uncertainties in our stellar and binary models propagate to uncertainties in whether a mass-transferring system experiences a \ac{CEE}. We compare the response of the radius of the donor to mass loss  to the response of the orbit to mass transfer to determine whether a binary experiences a \ac{CEE} (see Section \ref{subsec:underlyingPhysics}). This approach relies on determining the appropriate response of the donor to (adiabatic) mass loss, the amount of mass that the companion can accrete, and the specific angular momentum removed from the binary by the non-accreted mass; all of these quantities have uncertainties and are model-dependent. Other population synthesis codes directly use the mass ratio at \ac{RLOF} to determine whether the mass transfer will be stable (e.g. \citealt{hurley2002evolution}, \citealt{claeys2014theoretical}).

The evolution of a mass transferring system is non-trivial. One possibility is delayed dynamical instability, in which the donor experiences a prolonged mass transfer phase before it becomes dynamically unstable \citep{HjellmingWebbink1987,IvanovaTaam2004,Ge2010}. This can lead the donor to be significantly under-luminous at the moment when the \ac{CEE} begins, compared to its appearance at the onset of the mass transfer episode itself \citep{Podsiadlowski2002}. We report the properties at the onset of the \ac{RLOF} because we do not account for delayed dynamical instability.

The opposite is also possible, where initially unstable systems may reach a stable configuration after ejecting only a fraction of the common envelope. \cite{pavlovskii2016stability} found that some massive giant donors with stellar mass black-hole companions, which were previously expected to experience a \ac{CEE}, might experience stable mass transfer instead.

In general, the transition from stable to unstable mass transfer is not fully understood. The details of the threshold of stability are regulated by the hydrodynamics of the material lost from the donor and the thermal response of the donor. This confluence of physical processes will always be at play because, at the low mass exchange rates that mass loss is initiated at, both dynamical and thermal readjustments of the binary are taking place on competing timescales. Accessing this regime in three-dimensional hydrodynamic simulations is challenging, as discussed by \citet{Reichardt2019CEE} because near-stable configurations evolve over long timescales and are susceptible to small numerical perturbations. In the case of eccentric orbits, dynamical tides raised on the donor also play a significant role and the eventual orbital evolution is determined by a combination of tidal, thermal, and hydrodynamic evolution. 

\subsubsection{Tidal evolution}\label{subsec:tidalUncertainty}
Both the tidal circularisation timescales and the applicability of various types of tides are highly uncertain.  As discussed in Sections \ref{subsec:tides} and \ref{subsec:subpopulations}, we explore the impact of assumptions about the dominant tidal mechanism based on the evolutionary phase of the donor by considering the tidal circularisation timescale at the onset of \ac{RLOF}.  While this approach makes it possible to analyse the impact of different choices without re-analysing the full population, it does mean that tides are not self-consistently included throughout the evolution of the binary.

We also make a number of simplifying assumptions about the efficiency of tidal circularisation.  For example, we apply the equilibrium tide to convective-envelope donors regardless of the orbital eccentricity, although the perturbation-from-equilibrium approximation is unlikely to be valid for very eccentric binaries which are not pseudo-synchronised at periapsis.  We crudely approximate coefficients in the tidal circularisation timescale equations based on \cite{Rasio1996Tides}, \cite{hurley2002evolution} and references therein.

\cite{WitteSavonije1999b,WitteSavonije1999} discuss how resonance locking could enhance pre-\ac{RLOF} circularisation of a $10\ \rm{M_{\odot}}$ \ac{MS} star with a $1.4\ \rm{M_{\odot}}$ \ac{NS} companion. This system is similar to the phase immediately after the first \ac{SN} in Channel I. The timescales on which resonance locking occurs are typically a few million years and could lead to less eccentric orbits at the onset of \ac{RLOF}. It is uncertain how much of an effect resonance locking would have on a population of massive interacting binaries.

\subsubsection{Zero-eccentricity initial distribution}
\label{subsubsection:eccDist}
We assume that binaries are circular at birth. This assumption is justified for close binaries, which are tidally circularised at birth, but is not consistent with observations of wide binaries (see \citealt{Levato1987,Abt1990,sana2012binary,Kobulnicky:2014} as presented in Figure 3 of \citealt{MoeDiStefano2017}). 
Our goal is to be conservative when studying eccentricity at the onset of the \ac{CEE}. Thus all changes in eccentricity from an initially circular binary are due to the subsequent binary evolution. \cite{VignaGomez2018DNSs} showed that using a thermal eccentricity distribution at birth decreases the \ac{DNS} formation rate by about a half, but has no significant effect on the orbital properties of \acp{DNS}
(we follow the eccentricity distribution from \citealt{VignaGomez2018DNSs} to be able to make a more direct comparison with those results).
However, the eccentricity distribution at \ac{ZAMS} likely affects the eccentricity distribution  at the onset of \ac{RLOF}. 
This is particularly true for Channel II binaries, which enter the double-core \ac{CEE} without previous interactions or supernovae, thus retaining their birth eccentricity modulo tidal effects.  The current $e=0$ peak associated with these system (see Figure \ref{fig:orbitalProperties}) would be replaced by the birth eccentricity distribution.

\subsubsection{Massive binary stars}
In this work we focused on \acp{CEE} during the formation of \acp{DNS}. Similar evolutionary pathways are experienced by other massive stellar binaries, including progenitors of,  black hole - \ac{NS} or black hole - black hole systems \citep{dominik2012double,Kruckow2018ComBinE,Neijssel2019}. The impact of eccentric \ac{RLOF} is not in the exclusive interest of \ac{DCO} formation. The role of tidal evolution and dynamical instability is fundamental for massive stellar mergers \citep{Podsiadlowski+1992,justham2014luminous,VignaGomez2019HPPISN}.

\subsubsection{Dynamics}
We do not consider the impact of dynamical interactions on the formation of \acp{DNS}.  This could take the form of dynamically-induced mergers in dense stellar environments such as globular clusters (\citealt{AndrewsMandel:2019} but see, e.g., \citealt{Ye:2019}).  Meanwhile, Kozai-Lidov oscillations \citep{kozai1962secular,lidov1962evolution} in hierarchical triple systems can drive up the eccentricity of the inner binary (see \citealt{Naoz2015Review} for a review) and contribute to the formation of merging \acp{DNS} \citep{HamersThompson:2019}.  Both types of dynamical encounters can change the binary orbital evolution, including the eccentricity.

\subsubsection{\ac{DNS} merger rates}\label{subsubsec:rates}
The merger rate of \acp{DNS} was inferred to fall in the range 110--3840 Gpc$^3$ yr$^{-1}$ with 90\% confidence \citep{GWTC12019} based on a single detection of GW170817 with a flat-in-rate prior \citep{abbott2017gravitational}. 
The detection of GW190425 under the assumption of a \ac{DNS} progenitor updates the local \ac{DNS} merger rate to 250--2810 Gpc$^3$ yr$^{-1}$ \citep{abbott2020gw190425}.

\cite{VignaGomez2018DNSs} predicted a \ac{DNS} merger rate of $\approx 280$ Gpc$^{-3}$ yr$^{-1}$.
After several changes in COMPAS (see Section \ref{subsec:comparison}) the \ac{DNS} merger rate is now $\approx \MergingBNSRateGpcYear$ Gpc$^{-3}$ yr$^{-1}$. This is more in line with the reported rates from other rapid population synthesis codes \citep{Giacobbo2018MOBSE,GiacobboMapelli2019,Kruckow2018ComBinE}. \cite{chruslinska2018double} pointed out that all population synthesis models struggle to jointly predict the binary black-hole and \ac{DNS} merger rate, particularly under standard assumptions.
Most rapid population synthesis studies agree that predicted rates are difficult to reconcile with the \ac{DNS} merger rates inferred from gravitational-wave observations, particularly if future observations push these above $\gtrsim 1000\ \rm Gpc^{-3}\ yr^{-1}$. Some particular choices of physics increase rates up to a few hundreds of \acp{DNS} $\rm Gpc^{-3}\ yr^{-1}$, but the typical prediction is around a few tens of \acp{DNS} $\rm Gpc^{-3}\ yr^{-1}$. 
Recently, \cite{GiacobboMapelli2020} proposed that a revised natal kick prescription might prove important to reconcile double compact object merger rates, as suggested by \cite{chruslinska2018double} (see more about natal kick prescriptions in Section \ref{subsec:otherStudies}).

The impact of the power-law exponent ($\alpha_{\rm IMF}$) of the initial mass function on the \ac{DNS} merger rate has been previously discussed by \cite{deMinkBelczynski2015}.  They find that the rate can increase (decrease) by a factor of a few for plausible shallower (steeper) initial mass functions.   \citet[][see also \citealt{FarrMandel2018} and \citealt{Schneider2018response}]{Schneider2018} find that the initial mass function of young massive stars in the
30 Doradus region of the Large Magellanic Cloud may be shallower than the canonical \cite{salpeter1955luminosity} value.  While this level of fluctuation in the initial mass function may prove insufficient to resolve the \ac{DNS} rate discrepancy \citep{belczynski2018GW170817}, it may be one of the ingredients.

Observations of Galactic \acp{DNS} as radio pulsars point to a \ac{DNS} merger rate that peaks at a few tens of mergers per Myr in the Galaxy; e.g., $37_{-11}^{+24}$ Myr$^{-1}$ according to \citet{Pol:2020}.   This can be converted to a volumetric rate in units of Gpc$^{-3}$ yr$^{-1}$ by multiplying the rate per Myr per Milky Way equivalent galaxy by a factor of $\sim 10$ \citep{ratesdoc}.  The peak of the rate extrapolated from Galactic observations thus falls between the typical binary population synthesis predictions and the peak of the rate inferred from gravitational-wave observations.  However, this extrapolation does not account for differences between the Galaxy and other environments \citep{ratesdoc}, and all rate intervals are broad.

\citet{Lau2019LISASDNSs} use a synthesised \ac{DNS} population to predict that a four-year LISA mission  \citep{AmaroSeoane2017LISA,Baker2019LISA} will detect 35 Galactic \acp{DNS}. \cite{Andrews2019LISA} predict between 46 and 240 Galactic \acp{DNS} for the same mission, depending on the assumed physical assumptions. LISA \ac{DNS} observations will further constrain the \ac{DNS} formation and merger rates.

\subsubsection{Be X-ray Binaries}
The Be X-ray binary phase occurs in binaries consisting of a rapidly rotating Be star, a \ac{MS} B star with emission lines from a decretion disk, and a \ac{NS} which accretes from this disk. This phase can be one of the intermediate stages between \ac{ZAMS} and \ac{DNS} formation. 
We expect that most systems formed through Channel I experienced a Be X-ray binary configuration after the primary initiated the first mass transfer episode, spinning up the secondary, and exploded as a supernova (see Figure \ref{fig:channels}).

The SMC Be X-ray binary catalogue by \cite{CoeKirk2015} presents 69 systems. From those 69, 44 have an observed orbital period ($P_{\rm{orb}}<520$ days).
\cite{VinciguerraNeijssel2020} used COMPAS to study Be X-ray binaries at SMC metallicity ($Z\approx 0.0035$).
We focus on two particular variations from that study. Their \textit{default} model follows similar COMPAS settings as the ones in this paper, while the \textit{preferred} model allows for mass transfer to be more conservative than the default model, i.e. half of the mass transferred is accreted by the companion during the mass transfer episode.
The default model predicts $190\pm20$ Be X-ray binaries in the SMC; in this model, $\sim 59\%\ (46\%)$ of all (merging) \ac{DNS} experience a Be X-ray binary phase.
The preferred model predicts $80\pm10$ Be X-ray binaries in the SMC, with $\sim 96\%\ (98\%)$ of all (merging) \ac{DNS} experiencing a Be X-ray binary phase.
Both variations have a formation rate of merging \acp{DNS} of $\sim 10^{-5}\ \rm{M_{\odot}^{-1}}$.
\cite{VinciguerraNeijssel2020} show that the preferred model better represents observations of Be X-ray binaries in comparison to the default model. However, there is no significant difference in the predicted \ac{DNS} merger rate or the \ac{DNS} mass distribution between these two models.

\subsection{Comparison with other rapid population synthesis studies}
\label{subsec:otherStudies}
It is challenging to compare results between population synthesis studies.  The results depend on initial conditions, physical parameterisations and computational methods. Most population synthesis software differ in the implementation of at least one of these, and often in all. Some of the arguably most important physical interactions, including supernovae and mass transfer events, are treated differently between various research groups, codes and particular projects. 

In the context of supernovae, there are several common choices for remnant mass and natal kick prescriptions.  \texttt{StarTrack}, MOBSE and COMPAS usually follow \cite{fryer2012compact} for the remnant mass prescription of compact objects, using either the rapid or delayed variation, while \textsc{ComBinE} use their own \citep{Kruckow2018ComBinE}.
The natal kick distribution varies not only between codes, but also between papers. COMPAS initially followed \texttt{StarTrack} \citep[e.g.][]{belczynski2008compact,belczynski2018GW170817} with a no-kick model for \ac{ECSN} \citep{stevenson2017formation} and more recently changed to a low-kick distribution \citep[e.g.][]{VignaGomez2018DNSs}. MOBSE has explored the impact of variations in the \ac{CCSN} and \ac{ECSN} natal kicks \citep{GiacobboMapelli2018,GiacobboMapelli2019ECSN} and \textsc{ComBinE} use their own natal kick prescription \citep{Kruckow2018ComBinE}. COMPAS and \textsc{ComBinE} use a distinct prescription for \acp{USSN}; \texttt{StarTrack} initially did not separately account for them, but \cite{chruslinska2018double} followed the model from \cite{bray2016}, in which the natal kick is proportional to the ejecta mass. Similarly, MOBSE evolved from not explicitly accounting for \acp{USSN} \citep{GiacobboMapelli2018} to treating \acp{USSN} as low-kick \acp{CCSN} \citep{GiacobboMapelli2019ECSN}. \cite{GiacobboMapelli2020} estimate the natal kick using the ejecta mass, similarly to \cite{bray2016}.

In the context of mass transfer and \acp{CEE}, both crucial phases in \ac{DNS} formation, the main differences appear in the definition of dynamical instability and in the treatment of stable mass transfer and \acp{CEE}. 

For the definition of stability MOBSE follows BSE \citep{hurley2002evolution}. \texttt{StarTrack} uses adiabatic mass-loss indexes and a temperature threshold to identify stars with a fully-convective envelope. COMPAS generally follows BSE and \texttt{StarTrack} but with significant amount of changes (see \citealt{VignaGomez2018DNSs} and Appendix \ref{app:popSynth}). \textsc{ComBinE} use their own criteria based on a critical mass ratio \citep{Kruckow2018ComBinE}. 

COMPAS and MOBSE both follow BSE to solve for mass transfer episodes. In the default COMPAS models, non-accreted mass leaves the binary with the specific angular momentum of the accretor. 
\texttt{StarTrack} sets a fixed fraction of mass transfer as conservative, while any non-accreted mass leaves the system with the specific angular momentum of the orbit. \cite{Kruckow2018ComBinE} follows the approach from \cite{soberman1997stability}.

For the \ac{CEE}, COMPAS closely follows \texttt{StarTrack} as implemented in \cite{dominik2012double}, particularly in the estimation of the binding energy structure parameter $\lambda$ (see Section \ref{subsec:underlyingPhysics}). MOBSE follows \cite{Dewi2000CEE} and \cite{claeys2014theoretical} to determine this $\lambda$ parameter. Additionally, MOBSE frequently has highly-efficient envelope ejection by allowing $\alpha > 1$. Following \cite{Kruckow:2016tti}, \textsc{ComBinE} self-consistently uses their own stellar models to calculate the binding energy.

Most comparisons between population synthesis codes come from merger rates. However, the merger-rate density is not the only prediction from population synthesis, and the full properties of observed populations will place stronger constraints on the physics \citep{barrett2017accuracy, Kruckow2018ComBinE}. The population of luminous red novae which will become \acp{DNS} do not seem to be an unequivocal additional constraint (see Section \ref{subsec:LRNe}). Be X-ray binaries and short gamma-ray bursts must be included as additional constrains to the observed Galactic \ac{DNS} population and the growing catalogue of \ac{DNS} gravitational-wave sources to fully dissect the origin and formation of \acp{DNS}.

\section{Summary and Conclusions}
\label{sec:summaryAndConclusions}
We carried out a rapid population-synthesis study of a million massive binaries using COMPAS, finding \numberOfDNSsWithCEEs\ (unweighted) simulated systems which experience a \ac{CEE} and eventually become a \ac{DNS}. We present the key properties of the donor and binary star at the onset of the \ac{RLOF} phase leading to the \ac{CEE}. We provide an online catalogue of this synthesised population. 

Some of our main results are:
\begin{itemize}
    \item The \acp{CEE} that occur in \ac{DNS} progenitors can be broadly divided into two types (description in Sections \ref{subsec:formation} and \ref{subsec:subpopulations}). Channel I, which accounts for \fractionChannelI\ of all formed \acp{DNS}, involves a  post-\ac{MS} donor (the initially less massive star) with a \ac{NS} companion.  Channel II, which accounts for \fractionChannelII\ of all formed \acp{DNS}, involves two giant stars, with mass ratio close to unity, in a double-core common-envelope.
    \item Close to 10\% of the brightest luminous red nova transients, which have been previously associated with stellar mergers and common-envelope ejections, are predicted to occur during binary evolution that leads to \ac{DNS} formation. The progenitor of \LRNevent\ as reported in \cite{Blagorodnova2017LRN} is somewhat similar to the pre-CEE properties of DNS-forming systems (see Section \ref{subsec:LRNe} and Figure \ref{fig:bigHR}).
    \item We find that tidal circularisation timescales can be long compared to stellar radial growth timescales (see Figures \ref{fig:convectiveTidalTimescales} and \ref{fig:comparisonCircularisationTimescales}), especially for rapidly evolving \ac{HG} donors and/or donors with radiative envelopes experiencing only the less efficient dynamical tide rather than the more efficient equilibrium tide. This indicates that $\sim 20\%-70\%$ of binaries may not circularise prior to the onset of \acp{CEE} (see Sections \ref{subsec:TidalTimescales} and  \ref{subsec:subpopulations}). 
    This finding suggests that the ensuing common envelope phases in these binaries may be distinct from those that have been previously considered. Future work is needed to determine the implication of these differences for the predicted formation rate and properties of \acp{DNS}.
\end{itemize}

One of the main goals of this study is to constrain the parameter space of interest for detailed evolutionary studies of \acp{CEE}. We hope that the results presented in this catalogue can inform choices of initial conditions for detailed hydrodynamical simulations and lead to an improved understanding of the complexities of dynamically unstable mass transfer and the subsequent common-envelope phase. In particular, our present work highlights the roles of several uncertain processes that may be of crucial importance in \ac{DNS} formation:
\begin{enumerate}[label=(\roman*)]
    \item Tidal dissipation in pre-CEE binary evolution;
    \item Eccentric Roche lobe overflow; and
    \item The hydrodynamics of double-core CEEs.
\end{enumerate} 

\begin{acknowledgements}
We thank Poojan Agrawal, J.J. Eldridge, Mike Y.M. Lau, Orsola de Marco, Noam Soker, Simon Stevenson and Eric Thrane for discussions. We thank Teresa Rebagliato for the graphic representation of the formation channels.
A.V-G. acknowledges funding support from Consejo Nacional de Ciencia y Tecnolog\'ia (CONACYT) and support by the Danish National Research Foundation (DNRF132).
M.M. is grateful for support for this work provided by NASA through Einstein Postdoctoral Fellowship grant number PF6-170169 awarded by the Chandra X-ray Center, which is operated by the Smithsonian Astrophysical Observatory for NASA under contract NAS8-03060. This material is based upon work supported by the National Science Foundation under Grant No.~1909203.
S.dM. and S.J. acknowledge funding by the Netherlands Organization for Scientific Research (NWO) as part of the Vidi research program BinWaves with project number 639.042.728 and the European Union's Horizon 2020 research and innovation program from the European Research Council (ERC, grant agreement No.\ 715063).
I.M. is a recipient of the Australian Research Council Future Fellowship FT190100574.
\end{acknowledgements}

\newpage
\begin{appendix}
\section{Population synthesis: details of the COMPAS setup}\label{app:popSynth}
We present a list of the initial values and default settings used for this study in Table \ref{tab:population-synthesis-settings} in order to be able to emulate them with other population synthesis codes. References have been added where needed in order to justify our assumptions. Some of these assumptions are described in Section \ref{sec:methods}.

\begin{table*}
\caption{
Initial values and default settings of the population synthesis model simulations with COMPAS, including pertinent references.
}

\label{tab:population-synthesis-settings}
\centering
\resizebox{\textwidth}{!}{%
\centering
\begin{tabular}{lll}
\hline  \hline
Description / Name / Symbol   & Value / Range   & Note / Setting   \\ \hline  \hline
\multicolumn{3}{c}{Initial conditions}                                                                      \\ \hline
Initial mass $\monei$
& $[5, 100]$ \Msun
& \citet{kroupa2001variation} IMF  $\propto  {\monei}^{-\alpha}$  with $\alpha_{\rm{IMF}} = 2.3$ for stars above $1$\Msun \\
Initial mass ratio $\qi = \mtwoi / \monei $ 
& $[0.1, 1]$
& we assume a flat mass ratio distribution  $p(\qi) \propto  1$ \\
Initial separation $\ai$
& $[0.01, 1000]$ \AU & 
distributed flat-in-log $p(\ai) \propto 1 / {\ai}$ \\
\ac{SN} natal kick magnitude \vk                          							& $[0, \infty)$ \kms & 
drawn from Maxwellian distribution with standard deviation $\sigma$ \\
\ac{SN} natal kick polar angle $\thetak$
& $[0, \pi]$
& $p(\thetak) = \sin(\thetak)/2 $ \\
 \ac{SN} natal kick azimuthal angle $\phi_k$
 & $[0, 2\pi]$
 & uniform $p(\phi) = 1/ (2 \pi) $   \\
\ac{SN} mean anomaly of the orbit
& $[0, 2\pi]$
& uniform distributed  \\
Initial metallicity $\ensuremath{Z}$
& $Z_{\rm{\odot}}=0.0142$
& \cite{asplund2009chemical} \\
Initial orbital eccentricity $\ei$ 
& 0
& all binaries are assumed to be circular at birth  \\
\hline
\multicolumn{3}{c}{Settings} \\ \hline
Stellar winds for hydrogen-rich stars
& \citet{Belczynski2010a} & 
based on {\citet{Vink2000,Vink2001}}, including  luminous blue variable (LBV) wind mass loss with $f_{\rm{LBV}} = 1.5$. \\
Stellar winds for hydrogen-poor/helium stars & 
\citet{Belczynski2010b} & 
based on {\citet{HamannKoesterke1998}} and {\citealt{Vink2005}}. \\
Mass transfer stability criteria & $\zeta$-prescription & based on \citet[][]{VignaGomez2018DNSs} and references therein \\ 
Case BB mass transfer stability 
& always stable
& based on \cite{VignaGomez2018DNSs}        \\ 
CE prescription & 
$\alpha-\lambda$ & 
based on \citet{webbink1984double} and \citet{de1990common} \\
CE efficiency $\alpha$-parameter 
& 1.0 & \\
CE $\lambda$-parameter 
& $\lambda_{\rm{Nanjing}}=\lambda_{\rm{b}}$ 
& based on \citet{xu2010binding,XuLi2010Erratum} and \citet{dominik2012double}. Includes internal energy. \\
Hertzsprung gap (HG) donor in \ac{CE} 
& optimistic  
& defined in \citet{dominik2012double}:  HG donors survive a \ac{CE}  phase        \\
Core-collapse  \ac{SN} remnant mass prescription
& delayed
& from \cite{fryer2012compact} which has no lower black-hole mass gap \\
\ac{USSN}  remnant mass prescription
& delayed 
& from \cite{fryer2012compact} \\
ECSN remnant mass prescription 
& $m_{\rm{f}} = 1.26\Msun$ &
based on Eq.~8 in \citet{Timmes1996} \\
Core-collapse \ac{SN} velocity dispersion $\sigma_{\rm{rms}}^{\rm{1D}}$ 			& 265\kms & 
1D rms value based on \citet{hobbs2005statistical} \\
USSN and ECSN velocity dispersion $\sigma_{\rm{rms}}^{\rm{1D}}$ 
& 30\kms 
& 1D rms value based on e.g. \citet{pfahl2002population,podsiadlowski2004effects} \\
PISN / PPISN remnant mass prescription & NA & NA \\
\hline
\multicolumn{3}{c}{Simulation settings} \\ \hline
Total number of binaries sampled per metallicity $N_{Z}$ & 
$\approx 10^6$ & \\
Sampling method & 
\sc{STROOPWAFEL} & 
adaptive importance sampling from  \citet{broekgaarden2019stroopwafel} \\
Binary fraction & 
$f_{\rm{bin}} = 1$ &
broadly consistent with {\citet{sana2012binary}} \\
\hline \hline
\end{tabular}%
}
\end{table*}

\end{appendix}

\bibliographystyle{pasa-mnras}
\bibliography{cataloguePaper}

\end{document}